\documentclass[ 
reprint,
superscriptaddress,
amsmath,amssymb,
aps,
]{revtex4-2}

\usepackage{graphicx}
\usepackage{dcolumn}
\usepackage{comment}

\begin{document}

\preprint{APS/123-QED}

\title{Statistical properties of probabilistic context-sensitive grammars}

\author{Kai Nakaishi}
\affiliation{
    Graduate School of Arts and Sciences,
    The University of Tokyo,
    3-8-1 Komaba, Meguro-ku, Tokyo 153-8902, Japan
}
\author{Koji Hukushima}
\affiliation{
    Graduate School of Arts and Sciences,
    The University of Tokyo,
    3-8-1 Komaba, Meguro-ku, Tokyo 153-8902, Japan
}
\affiliation{
    Komaba Institute for Science, The
    University of Tokyo, 3-8-1 Komaba, Meguro-ku, Tokyo 153-8902, Japan
}

\begin{abstract}
    Probabilistic context-free grammars (PCFGs), which are commonly used to generate trees randomly, have been well analyzed theoretically, leading to applications in various domains. Despite their utility, the distributions that the grammar can express are limited to those in which the distribution of a subtree depends only on its root and not on its context. This limitation presents a challenge for modeling various real-world phenomena, such as natural languages. To overcome this limitation, a probabilistic context-sensitive grammar (PCSG) is introduced, where the distribution of a subtree
    depends on its context. 
    Numerical analysis of a PCSG reveals that the distribution of a symbol does not constitute a qualitative difference from that in the context-free case, but mutual information does. 
    Furthermore, a novel metric introduced to directly quantify the breaking of this limitation detects a distinct difference between PCFGs and PCSGs.
    This metric, applicable to an arbitrary distribution of a tree, allows for further investigation and characterization of various tree structures that PCFGs cannot express.
\end{abstract}

\maketitle

\section{Introduction}

Hierarchical structures underlie many real-world phenomena, including natural languages. A context-free grammar (CFG), a fundamental concept in formal language theory, was originally introduced to analyze hierarchical syntactic structures in natural languages \cite{Chomsky_SS}. Furthermore, it provides a basis for describing more general hierarchical structures that are not limited to natural languages. 
A CFG, defined by a set of production rules, generates strings with trees in a formal way. The strings correspond to sentences, whereas the trees describe the hierarchical syntactic structures behind the sentences.
A probabilistic extension of a CFG, known as a probabilistic context-free grammar (PCFG) or stochastic context-free grammar \cite{Jelinek1992}, introduces probabilities into the production rules. 
According to the rules, this model generates trees in a probabilistic manner. 
This probabilistic grammar has been used to model syntactic structures of a natural language \cite{Charniak1997} or a programming language \cite{Ellis2015}, and to study many other phenomena with tree or hierarchical structures in fields such as music \cite{Worth2005, Gilbert2007}, human cognition \cite{Tano2020}, long-short-term-memory network \cite{Lin2017}, RNA \cite{Knudsen1999}, cosmic inflation \cite{Harlow2012}, or a more abstract model \cite{Li1989open, Li1991, Lieck2021}.
Additionally, other frameworks are closely related to a PCFG, including a branching process and a Lindenmayer system (or L-system) \cite{Lindenmayer1968I, Lindenmayer1968II, Herman1974}.

An essential property of a PCFG is that the distribution of a subtree depends only on its root, not on the context, which we will designate as context-free independence.
This property allows an exact mathematical analysis of the statistical properties of PCFGs. Indeed, earlier studies have analyzed and resolved various aspects of PCFGs, including the probability of symbol occurrence \cite{Nakaishi22}, the correlation function \cite{Li1989open, Li1991}, mutual information between nodes \cite{Lin2017}, the mean sentence length \cite{Booth1973}, entropy \cite{Miller1992, Chi1999}, branching rates \cite{Miller1992, Chi1999}, tree size \cite{Chi1999}, and the conditions for sentence generation to terminate with probability $1$ \cite{Booth1973, Esparza2013}. At the same time, this property is too strict to impose on real-world phenomena. Particularly, it is well known in linguistics that some languages in the real world cannot be described using a CFG because of its inability to represent cross-serial dependencies \cite{Shieber85, Culy85}.
In natural language processing, empirical evidence suggests that a naive parser relying on a PCFG is insufficient for inferring syntactic structures \cite{Charniak1997}. Moreover, certain parsers that relax context-free independence in technical manners can express more complex distributions and can achieve higher accuracy \cite{Johnson2006, ODonnell2009}. Outside of language-related domains, the possibility that introducing context sensitivity is useful for describing music is also discussed \cite{Worth2005}.
Therefore, the distributions that a PCFG can express are regarded as severely limited.

To understand more realistic phenomena with hierarchical structures, it is necessary to introduce and analyze a model that captures the distribution of a tree beyond context-free independence.
For this purpose, one can naturally consider context-sensitive grammars (CSGs) \cite{Chomsky_three}, which form the class one level higher than CFGs in the hierarchy of expressive power: the so-called Chomsky hierarchy. Similarly to a PCFG, a probabilistic context-sensitive grammar (PCSG) can be 
formulated as a probabilistic extension of a CSG.
A PCSG explicitly relaxes context-free independence. Consequently, the theoretical analyses developed for a PCFG are not applicable to a PCSG. The statistical properties of a PCSG have only rarely been analyzed, either theoretically or numerically.

To address this point, for this work, we defined a simple PCSG and investigated its statistical properties by numerical simulations systematically, mainly examining whether a qualitative difference from a PCFG exists, or not. To be more precise, we implemented a PCSG to measure the distribution of a symbol, mutual information between two nodes, and mutual information between two pairs of children of nodes on which the symbols are fixed. Here, we present a comparison of the observed similarities and differences between PCFG and PCSG: No qualitative difference was found in the distribution of a symbol between a PCSG and a PCFG.This result suggests that the properties observed in PCFGs are likely to be preserved in PCSGs. Given that the absence of a singularity in the distribution in an ensemble of PCFGs has been proven \cite{Nakaishi22}, it is reasonable to infer that PCSGs would not exhibit the singularity, similarly to PCFGs.
This singularity is relevant for the discussion on a phase transition in the random language model (RLM) \cite{DeGiuli19a, DeGiuli19b}, which might be analogous to discontinuity in human language acquisition according to earlier research.

However, the behaviors of mutual information between two nodes differ between a PCFG and a PCSG. 
The mutual information in a PCFG decays exponentially with the distance between two nodes, i.e., the path length in a tree graph. In contrast, in a PCSG, the mutual information decays exponentially with the effective distance, which is defined by considering the effect of context sensitivity.

Additionally, a more pronounced difference concerns the mutual information between pairs of children of symbol-fixed nodes. 
This novel metric, proposed in this research, quantifies context-free independence breaking. From a theoretical physics perspective, this metric represents the degree to which the network of interactions deviates from a tree structure. Linguistically, it represents the strength of mutual dependence between the structures of two constituents or phrases of given types. This metric not only detects whether context-free independence is broken; it also quantifies where and how strongly the breaking occurs.  In a PCFG, the context-free independence breaking is always zero. By contrast, in a PCSG, it is positive and decays similarly to the mutual information between nodes. As a result, the most striking difference between a PCFG and a PCSG is in this metric. This quantification is intuitive and is definable for any distribution of a tree. Measuring this metric in other mathematical models or real-world phenomena will help deepen the understanding of them by investigating how their behavior differs from that of a PCFG.

Here, we provide a brief summary of the main contributions made in this paper. Our first main contribution is the systematic investigation of a PCSG, which is a simple model for generating hierarchical structures beyond those produced by PCFGs. A key distinction between a PCFG and a PCSG is in the distance that determines the decay of mutual information. Second, we propose a novel metric for the context-free independence breaking, which has not been quantified previously. This metric allows for further quantitative investigation of various hierarchical structures that violate the context-free independence. Our results show that this metric decays exponentially for a PCSG while it remains zero for a PCFG, demonstrating the usefulness of this metric.

This paper is structured as follows: The models, a PCFG and a PCSG, are introduced in Sec.~\ref{sec_model}. The analysis of the distribution of a symbol in a PCSG and the argument about the phase transition in the RLM are presented in Sec.~\ref{sec_distribution}. Then, in Sec.~\ref{sec_mutual}, a numerical analysis of the mutual information between two nodes is presented, including the definition of the effective distance. The introduction and analysis of the quantification of the context-free independence breaking are given in Sec.~\ref{sec_violation}. Finally, we summarize the results and briefly discuss future works in the last section.

\section{Model}
\label{sec_model}

\subsection{Probabilistic context-free grammar}

In formal language theory \cite{Chomsky_three}, a grammar $G$ consists of a vocabulary $V$ and a finite set $R$ of rules.
A vocabulary $V$, a finite set of symbols, is divided into nonterminal symbols $A, B, \cdots \in V_N$ and terminal symbols $a, b, \cdots \in V_T$.
Each rule in $R$ is of the form $\varphi \to \psi$, meaning that a finite string $\varphi$ in $V$ is rewritten as another finite string $\psi$. Also, the left-hand side $\varphi$ of the rule must include at least one nonterminal symbol.
The grammar $G$ generates a sentence by the following process: Initially, a special symbol $S \in V_N$, called the starting symbol, is given.
Next, $S$ is rewritten by a rule $S \to \varphi$.
When a substring $\psi$ of $\varphi$ includes a nonterminal symbol, $\varphi$ can be rewritten by replacing $\psi$ with another string $\omega$ according to a rule $\psi \to \omega$.
This process is repeated.
Finally, if the string has no nonterminal symbol, it can no longer be rewritten by any rule.
The final string is called a sentence. The whole process of generating a sentence is called a derivation.
The set of sentences generated using a grammar $G$ is a language of $G$.
The importance of the finiteness of symbols and rules is noteworthy. If infinite sets $V$ and $R$ are allowed, then it becomes trivially possible to construct a grammar that generates an arbitrary language by introducing a symbol $A$ and a rule $A \to \varphi$ for each sentence $\varphi$ in the language. 
The infinite number of symbols or rules would make the concept of characterizing and classifying languages in terms of grammars irrelevant.

\begin{figure*}
    \centering
    \includegraphics[width = .8 \linewidth]{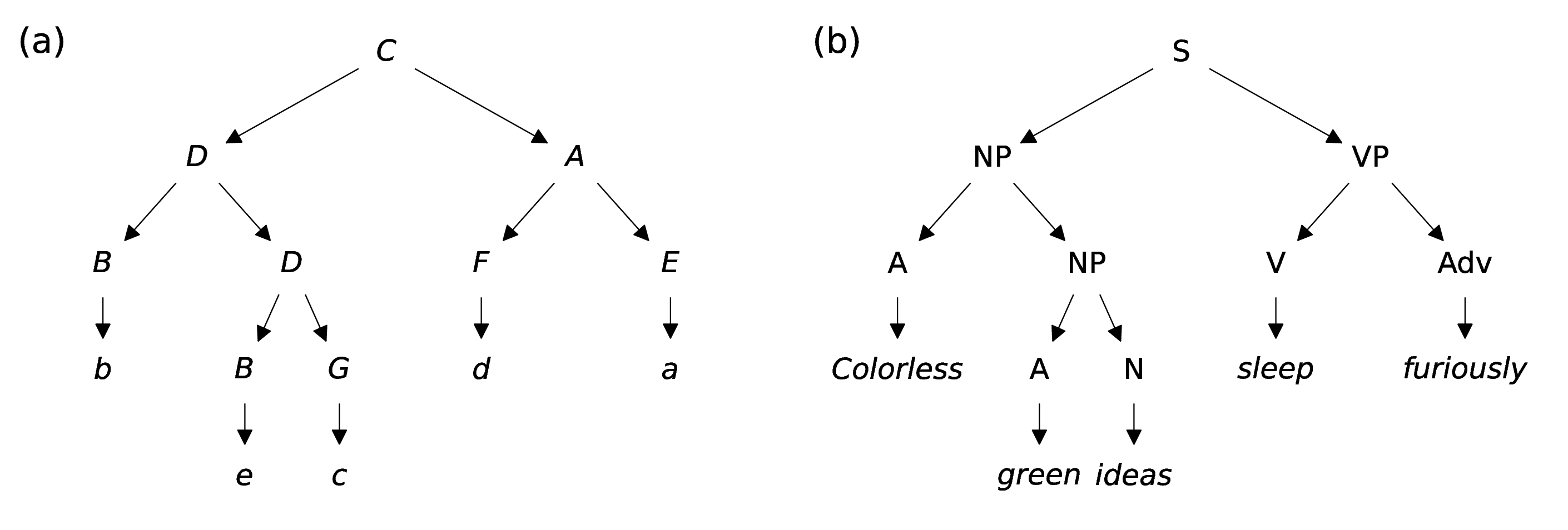}
    \caption{\label{fig_tree} (a) Example of a derivation generated using a CFG in CNF. A node with its children means that the node is rewritten as the children. In this example, the initial string is $C$. Applying the first rule $C \to DA$, the string becomes $DA$. The next rule $D \to BD$ (or $A \to FE$) rewrites the string as $BDA$ (or $DFE$). The remainder of the derivation is similar. The final string, i.e., the sentence, is $becda$.
        (b) Syntactic structure behind the sentence {\it Colorless green ideas sleep furiously} in terms of immediate constituent analysis. This diagram means, for instance, that the noun phrase (NP) {\it green ideas} consists of the adjective (A) {\it green} and the noun (N) {\it ideas}. Roughly speaking, a nonterminal symbol in a CFG corresponds to a constituent in syntax; a terminal symbol corresponds to a word.}
\end{figure*}

A grammar $G$ is a CFG \cite{Chomsky_SS} if every rule of $G$ is of the form $A \to \varphi$ with $A$ being nonterminal.
The derivation in a CFG can be represented as a tree, which is analogous to the syntactic structure of a sentence in a natural language analyzed by immediate constituent analysis, as shown in Fig.~\ref{fig_tree}.
In fact, any CFG can be transformed to an equivalent CFG where every rule is of the form $A \to BC$ or $A \to a$ for $A, B, C \in V_N$ and $a \in V_T$, ensuring that the generated language remains unchanged.
This transformed form is referred to as the Chomsky normal form (CNF)\cite{Hopcroft07}.

A PCFG \cite{Jelinek1992} is a probabilistic version of a CFG.
It is introduced by assigning a probabilistic weight $M_{A \to \varphi}$ to each CFG rule $A \to \varphi$, meaning that a nonterminal symbol $A$ is rewritten as $\varphi$ with probability $M_{A \to \varphi}$.
The PCFG specified by the set of weights $M_{A \to \varphi}$ determines the probability of a derivation, which is the product of the weights of all rules applied in the derivation.
If we adopt the idea of simplifying a speaker or a group of speakers of a language as an agent that generates strings with syntactic structures probabilistically, then a PCFG can be a simple mathematical model for a language.
Indeed, a PCFG has been used for modeling a natural language \cite{Charniak1997} and a programming language \cite{Ellis2015}.
In addition, because a PCFG can be regarded as a simple mathematical model for randomly generating trees or hierarchical structures, many studies have used it as a model not only for a natural or formal language but also for other phenomena \cite{Worth2005, Gilbert2007, Tano2020, Lin2017, Knudsen1999, Li1989open, Li1991, Lieck2021}. A PCFG also has a close relation to other physical and mathematical frameworks \cite{Lindenmayer1968I, Lindenmayer1968II, Herman1974}.

By definition, the distribution of a subtree in a PCFG depends only on the root. It is unaffected by the context, i.e., the neighboring symbols of the root.
Because of this context-free independence, 
many properties of a PCFG can be analyzed theoretically. For instance, the distribution of a symbol or the joint distribution of several symbols at arbitrary nodes can be computed recursively from the root of the entire tree, similarly to a Markov chain. Indeed, many earlier studies have analyzed properties of a PCFG theoretically and exactly \cite{Nakaishi22, Li1989open, Li1991, Lin2017, Booth1973, Miller1992, Chi1999, Chi1999, Booth1973, Esparza2013}.
The context-free independence allows for the theoretical analysis of various properties of PCFGs, but it also severely restricts the range of distributions that a PCFG can express.
In general, it is not reasonable to expect a natural phenomenon to satisfy such a restriction.
Linguistically, for instance, some real-world languages cannot be described by CFGs \cite{Shieber85, Culy85}. Moreover, natural language processing researchers have found it necessary, empirically, to relax the context-free independence for modeling the syntactic structures of natural languages \cite{Johnson2006, ODonnell2009}.
However, no report of the relevant literature describes a systematic investigation of a simple mathematical model that goes beyond the independence or a quantitative analysis of the degree to which context-free independence is broken in any model or phenomenon.
This need for study prompts us to consider such a model and to quantify how far the model is from the independence.

\subsection{Probabilistic context-sensitive grammar}

A model introduced by allowing each rule in a CFG to refer to the context is a CSG, which has one level higher expressive power than a CFG in formal language theory \cite{Chomsky_three}.
In a CSG, a rule is of the form $\varphi A \psi \to \varphi \omega \psi$. In other words, the result $\omega$ of rewriting $A$ can depend on the substrings $\varphi$ and $\psi$ next to $A$, i.e., the context of $A$.
The class of languages generated by CSGs is believed to be larger than the class of possible natural languages \cite{jager2012formal}.
Additionally, we can naturally define a probabilistic version of a CSG, namely, a PCSG, by assigning a probabilistic weight to each rule, similar to the introduction of a PCFG.
A PCSG relaxes the context-free independence, 
meaning that
the distribution of a subtree in a PCSG depends not only on its root but also on the context.
The theoretical analyses of a PCFG described above \cite{Nakaishi22, Li1989open, Li1991, Lin2017, Booth1973, Miller1992, Chi1999, Chi1999, Booth1973, Esparza2013}, all of which impose the independence, are not applicable to a PCSG.
Consequently, the behavior of a PCSG and its characteristics, such as which of its properties are similar to or different from those of a PCFG, are unknown.

The class of all possible grammars defined as a probabilistic extension of a CSG is too large and complicated to analyze. We focus, therefore, on a simpler model within a CSG to examine its behavior. First, we consider a CSG with a vocabulary consisting of binary nonterminal symbols, $V_N = \{ 0, 1 \}$. We do not consider terminal symbols.
In the following, a symbol simply means a nonterminal symbol unless otherwise noted.
Additionally, we restrict rules to the form of $A \to BC$ or $L A R \to L BC R$.
The former is a nonterminal rule of a CFG in CNF, whereas the latter is a CSG rule with context sensitivity that refers only to the two symbols next to the rewritten symbol.
Consequently, the cause of the difference between our model and the binary CFG or PCFG in CNF is, in essence, the context sensitivity to $L$ and $R$.
In our notation, $A$, $B$, and $C$ represent symbols, whereas $L$ and $R$ can be symbols or nulls $\lambda$.
For example, if the rule $\lambda 0 1 \to \lambda 111$ is applied to the leftmost 0 in the string 0110, then the string turns to 11110.

Our PCSG is defined as the probabilistic extension of this CSG. The probabilistic weight $M_{ABC}^{\mathrm{CF}}$ is assigned to each CFG rule $A \to BC$, and $M_{L A R, BC}^{\mathrm{CS}}$ to each CSG rule $LAR \to LBCR$. Next, we introduce the probability $q$ that a CSG rule is chosen, to control the degree of context sensitivity. More precisely, symbol $A$ in the context $LAR$ is rewritten as $BC$ by a CFG rule $A \to BC$ with probability $(1 - q) M_{ABC}^{\mathrm{CF}}$, or as $DE$ by a CSG rule $LAR \to LDER$ with probability $q M_{L A R, DE}^{\mathrm{CS}}$. A PCSG with $q = 0$ is a PCFG. Additionally, we must determine the order in which rules are applied to a string because, in a PCSG, unlike a PCFG, a derivation depends on the order. For this study, we 
choose to apply rules in a uniformly random manner as a neutral alternative. If the length of a present string is $l$, we first generate a random permutation $\tau$ of $\{0, \cdots, l-1 \}$ according to a uniform distribution, and then apply rules to the symbols sequentially, from the $\tau (0)$-th one to the $\tau(l-1)$-th one. After all symbols of the preceding string are rewritten, the length becomes $2 l$. The whole procedure to generate a tree is as follows: The first step in the derivation is to choose a symbol from a uniform distribution over $V_N$. Subsequently, a string is rewritten recursively. For each step, the order of application of rules and each rewriting are determined randomly in the manner we explained above.
Because no terminal symbol exists in this setting, a rule can always be applied to the string no matter how many steps the derivation goes through.
Consequently, we stop the process when the step is repeated $D$ times, which is a value determined in advance \footnote{Although the process of rewriting a symbol as a function of the symbol and its neighbors is similar to that of an elementary cellular automaton, our model has several features. In our model, rewriting operations are asynchronous and random. Furthermore, the number of cells or symbols grows exponentially because a single symbol becomes two symbols. A cellular automaton with asynchronous and random updates is called an asynchronous cellular automaton \cite{Fates2018}. Our PCSG can be regarded as a modified version of an asynchronous cellular automaton such that the system size grows.}.

Although the discussion in the remainder of this paper is based on the above setting, we have found that the properties of a PCSG remain qualitatively unchanged under alternative settings.
For example, the model exhibits similar behavior when each rule refers to two left neighbors and two right neighbors, or when symbols are rewritten in a different order, such as left-to-right or inside-to-outside.

This type of PCSG is specified by the probability $q$ and the weights $M = \left( M^{\mathrm{CF}}, M^{\mathrm{CS}} \right)$, where $M^{\mathrm{CF}} = \{ M^{\mathrm{CF}}_{ABC} \}_{ABC}$ and $M^{\mathrm{CS}} = \{ M_{LAR, BC}^{\mathrm{CS}} \}_{LAR, BC}$.
The probabilistic weights are sampled according to the lognormal distributions with normalization conditions
\begin{equation*}
    \begin{split}
        M^{\mathrm{CF}}_{ABC}
         & = \frac{
        \tilde{M}^{\mathrm{CF}}_{ABC}
        }{
        \sum_{B', C'} \tilde{M}^{\mathrm{CF}}_{AB'C'}
        },                      \\
        P \left(
        \tilde{M}^{\mathrm{CF}}_{ABC}
        \right)
         & \propto \mathrm{e}^{
        - \epsilon
        \ln^2 \tilde{M}^{\mathrm{CF}}_{ABC}
        },                      \\
        M^{\mathrm{CS}}_{LAR, BC}
         & = \frac{
        \tilde{M}^{\mathrm{CS}}_{LAR, BC}
        }{
        \sum_{B', C'}
        \tilde{M}^{\mathrm{CS}}_{LAR, B'C'}
        },                      \\
        P \left(
        \tilde{M}^{\mathrm{CS}}_{LAR, BC}
        \right)
         & \propto \mathrm{e}^{
        - \epsilon
        \ln^2 \tilde{M}^{\mathrm{CS}}_{LAR, BC}
        }.
    \end{split}
\end{equation*}
Therein, $\epsilon$ is the parameter used to control the width of the lognormal distributions.

For this study, we are interested in how the introduction of context sensitivity affects the statistical properties of PCFGs. Specifically, we implement PCSGs and conduct numerical analyses of three statistical quantities. The first involves the distribution of a symbol at a node, analogous to magnetization in a spin model. This quantity is related to the phase transition in the RLM \cite{DeGiuli19a, DeGiuli19b}.
The second specifically examines the mutual information between two nodes, which is associated with a two-point correlation.
Finally, we introduce the mutual information between the children of two symbol-fixed nodes.
This metric, which is zero for $q = 0$ by definition, reflects how strongly the independence is broken.

\section{Distribution of a Symbol}
\label{sec_distribution}

\subsection{Distribution of a symbol}

Primary emphasis should be on the distribution of a symbol on a single node.
We denote the probability that symbol $A$ occurs on node $i$ as
\begin{equation*}
    \pi_{A, i} \left( q, M \right)
    \equiv \left\langle
        \delta_{A, \sigma_i}
    \right\rangle_{q, M},
\end{equation*}
where $\sigma_i$ is a symbol on node $i$, and $\left\langle\cdots\right\rangle_{q, M}$ represents the average over trees under a PCSG with parameters $(q, M)$.
This quantity corresponds to the magnetization in the Potts spin model \cite{Wu1982}, where each site $i$ has a spin $\sigma_i$, and the magnetization along the direction $A$ is defined by the ratio of sites with $\sigma_i = A$.
In the case of $q = 0$, i.e., a PCFG, the context-free independence enables us to apply the concept of Markov chains. Because of this, the probability $\pi_{A, i}$ can be computed.
If node $i$ is the left child of node $j$, then $\pi_{B,i} = \sum_A \left( \sum_{C} M^{\mathrm{CF}}_{ABC} \right) \pi_{A, j}$.
If node $i$ is the right child, then it is the same except that $\pi_{B, i}$ and $\sum_C$ are replaced, respectively, with $\pi_{C,i}$ and $\sum_B$.
However, this no longer holds in the case of $q > 0$ because of the broken independence.

\begin{figure*}
    \centering
    \includegraphics[width = 1 \linewidth]{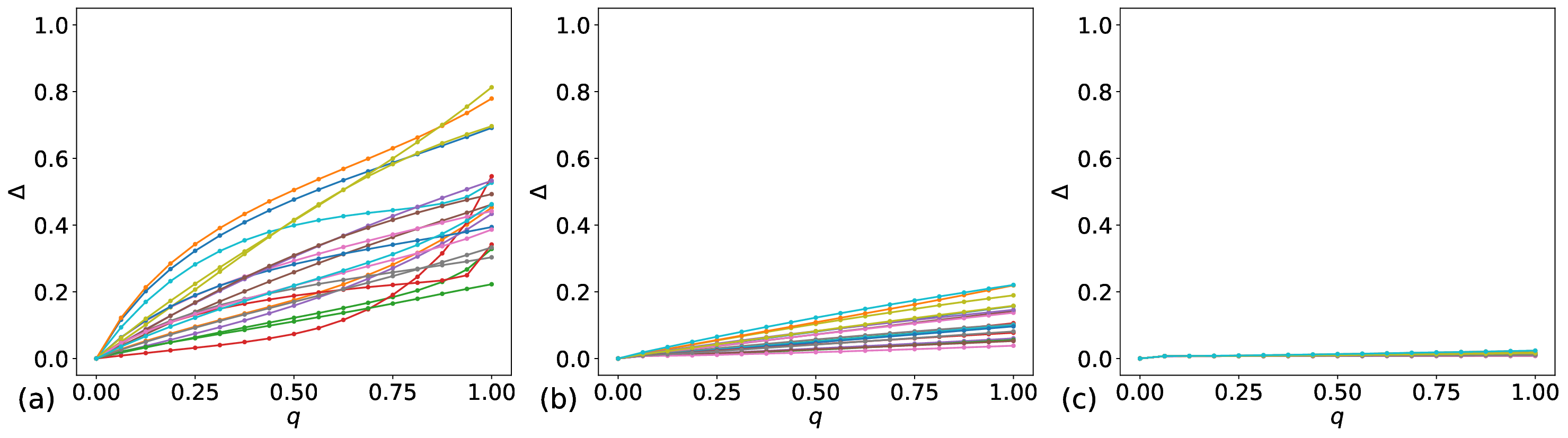}
    \caption{\label{fig_DP} Differences $\Delta (q, M)$ between $\pi_{A, i}$ with $q = 0$ and that with $q > 0$ as functions of $q$, computed from $10^4$ sampled trees of depth 10. Different colors represent different $M$'s. Panels (a), (b), and (c), respectively, present results for $M$'s generated from the lognormal distribution with $\epsilon = 10^{-2}$, $10^0$, and $10^2$.
    }
\end{figure*}

To see the degree to which the distribution of a symbol changes with the context sensitivity, we measured the Euclidean distance $\Delta$ between $\{ \pi_{A,i} \}_{A, i}$ with $q = 0$ and that with $q > 0$, expressed as
\begin{equation}
    \Delta (D, q, M)
    \equiv \sqrt{
        \frac{
            \sum_{i, A} \left(
            \pi_{A, i} (q, M)
            - \pi_{A, i} (0, M)
            \right)^2
        }{2 \left(2^{D + 1} - 1\right)}
    } .
\end{equation}
Figure~\ref{fig_DP} presents the distances $\Delta$ as functions of $q$ for $\epsilon = 10^{-2}, 10^0$ and $10^{2}$.
We sampled 20 $M$s for each $\epsilon$, and $10^4$ complete trees for each PCSG, with depth $D$ of a tree set to 10.
These figures show that $\Delta$ increases monotonically and continuously for any $M$.
It can also be observed that the increase is slower with larger $\epsilon$. If $\epsilon$ is larger, most of the generated $M^{\mathrm{CF}}_{ABC}$s and $M^{\mathrm{CS}}_{LAR,BC}$s are near $1/2^2$. As a result, $\pi_{A, i}$s are near $1/2$ for any $A$ and $i$ with most $M$s. This fact leads to the slower increase.
This behavior of $\Delta$ implies that the context sensitivity drives $\{ \pi_{A,i} (q, M) \}_{A, i}$ farther away, monotonically and continuously, from that for $q = 0$, and that no singularity occurs at any point in $0 < q < 1$.
It is noteworthy that the context sensitivity is not the only factor that contributes to this behavior, at least qualitatively.
Suppose we interpolate between a PCFG $M^{\mathrm{CF}}$ and another independently generated PCFG ${M^{\mathrm{CF}}}'$, instead of an $M^{\mathrm{CS}}$.
Even in this case, $\Delta$ will grow similarly with $q$.
It is not possible to see any qualitative difference between a PCFG and a PCSG in terms of the distribution of a symbol.

The observations presented here are for finite trees.
However, for most of the 20 $M$s, $\Delta$ with $D = 10$ seems to converge almost to that in the limit $D \to \infty$.
Consequently, it is unlikely that $\Delta$ has a singularity, even in the limit of infinite trees.
Supplemental Material 
provides numerical observations of how $\Delta$ converges as $D$ increases.

\subsection{Order parameter for the random language model}

In our case, because the tree topology is always the same, the mean ratio $\pi_A$ of symbol $A$ in a whole tree is the average of $\pi_{A,i}$ over nodes $i$. We denote it as
\begin{equation*}
    \pi_A \left( D, q, M \right)
    \equiv \frac{
        \sum_i \pi_{A,i} (q, M)
    }{
        2^{D + 1} - 1
    }.
\end{equation*}
The probability density of $\pi_{A}$ attributable to the randomness of $M$, defined as
\begin{equation*}
    P ( \pi_A | D, q, \epsilon )
    \equiv \int \mathrm{d} M P (M) \delta \left(
        \pi_A - \pi_A \left( q, M \right)
    \right),
\end{equation*}
plays a crucially important role in the discussion of the phase transition in the RLM, proposed in \cite{DeGiuli19a, DeGiuli19b}.
The RLM is defined as an ensemble of PCFGs generated according to the lognormal distribution, which is equivalent to the $q = 0$ case in our model.
An earlier study investigated the possibility of a phase transition characterized by the singularity of an \textit{order parameter} as the parameter $\epsilon$ varies.
This earlier study suggested that the phase transition can be interpreted as a possible discontinuity in human language acquisition.
However, recent findings in \cite{Nakaishi22} have revealed that the singularity of their order parameter, if any, is reduced to that of the probability density of $\pi_A$ and that the probability density is an analytic function of $\epsilon$ with finite vocabulary.
In other words, the phase transition does not exist as long as the number of types of symbols is finite.
This conclusion holds true for any analytic distribution of $M$, irrespective of whether it follows the lognormal distribution or whether the sizes of trees are finite or infinite.
Because the proof relies on the assumption of context-free independence, it cannot be extended to a context-sensitive case with $q > 0$.
Therefore, whether a phase transition exists in the context-sensitive RLM remains a non-trivial question.

To investigate whether the distribution of $\pi_A$ in the context-sensitive RLM has a singularity, we measured the Binder parameter of $\pi_A$, defined as
\begin{equation*}
    U ( D, q, \epsilon )
    \equiv 1 - \frac{
        \left[ ( \Delta \pi_A )^4 \right]_{\epsilon}
    }{
        3 \left[ ( \Delta \pi_A )^2 \right]_{\epsilon}^2
    },
\end{equation*}
where $\Delta \pi_A \equiv \pi_A - 1/2$ and $[ \cdots ]_{\epsilon}$ means the average over $M$s according to the lognormal distribution determined by $\epsilon$.
This parameter has been used to detect the transition in various statistical-mechanical models numerically \cite{Binder81, Binder84}.
This parameter is zero when $\pi_A$ follows a Gaussian distribution and nonzero when the distribution of $\pi_A$ is multimodal or non-Gaussian.
To compute the Binder parameter, we sampled $10^4$ $M$s for each $\epsilon$ and $10^3$ trees for each $M$.
Error bars were computed using the bootstrap method \cite{Efron1993,Young2012} with $10^2$ bootstrap sets.

\begin{figure*}[t]
    \centering
    \includegraphics[width = .8 \linewidth]{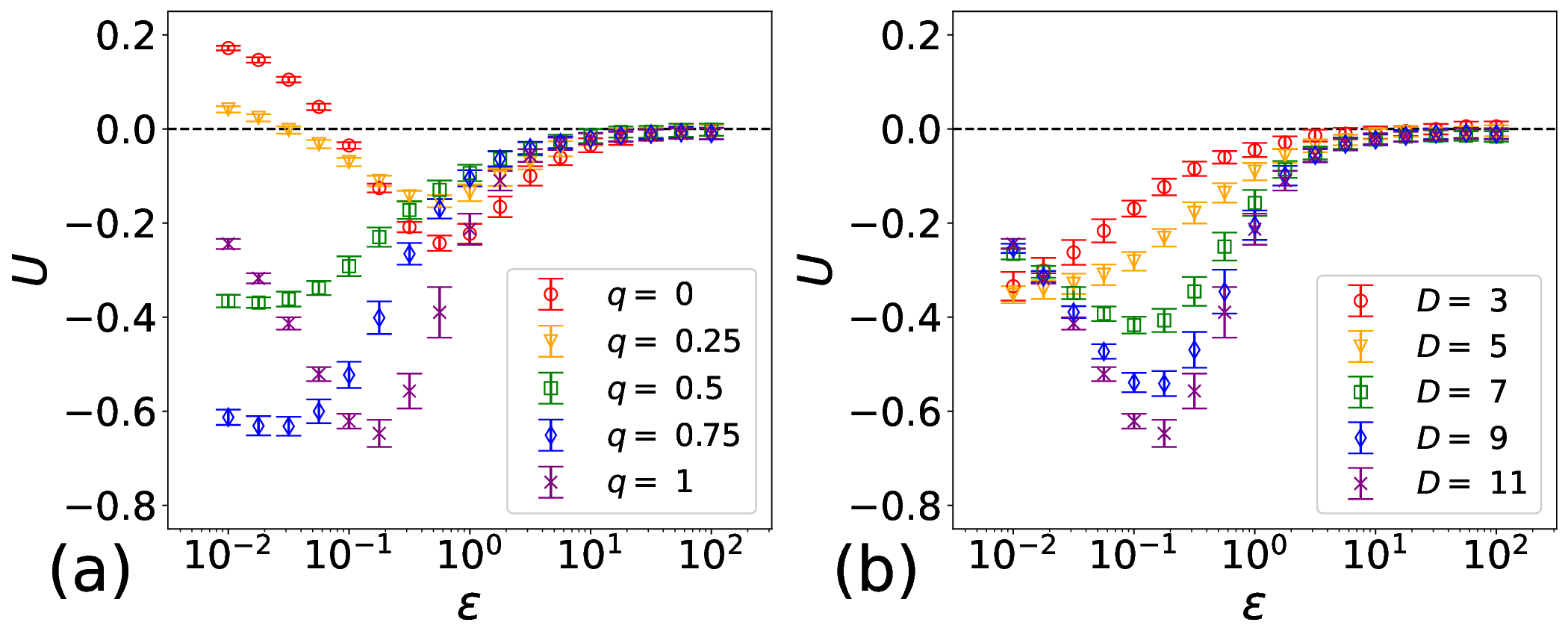}
    \caption{\label{fig_pi_Binder}
        Binder parameter $U$ of the mean ratio $\pi_A$ of symbol $A$ as a function of the parameter $\epsilon$
        (a) for depth $D = 11$ and context sensitivity $q = 0, 0.25, 0.5, 0.75$, or $1$, and (b) for $D = 3, 5, 7, 9$, or $11$ with $q = 1$.}
\end{figure*}

Figure~\ref{fig_pi_Binder}(a) shows the result obtained when the tree depth is fixed at $D = 11$ and the context sensitivity is $q = 0, 0.25, 0.5, 0.75$, and $1$.
From these findings, the Binder parameter seems to change analytically, but it changes more dramatically if the context sensitivity is larger.
Consequently, if the singularity exists, it might occur for $q = 1$. 
We also computed the Binder parameters for $q = 1$ while varying the depth $D$ of a tree, the result of which is shown in Fig.~\ref{fig_pi_Binder}(b).
For all previously known cases of phase transitions detected by this parameter, a discontinuous jump from zero to non-zero is found at the transition temperature in the thermodynamic limit.
However, it is unlikely that such a transition occurs for the limit $D \to \infty$ because the Binder parameter for large $\epsilon$ becomes farther away from zero as $D$ increases. Note that we do not rule out the possibility of another phase transition detected by other methods, which remains an open problem.

\section{Mutual Information Between Two Nodes}
\label{sec_mutual}

\begin{figure}
    \centering
    \includegraphics[width = 1\linewidth]{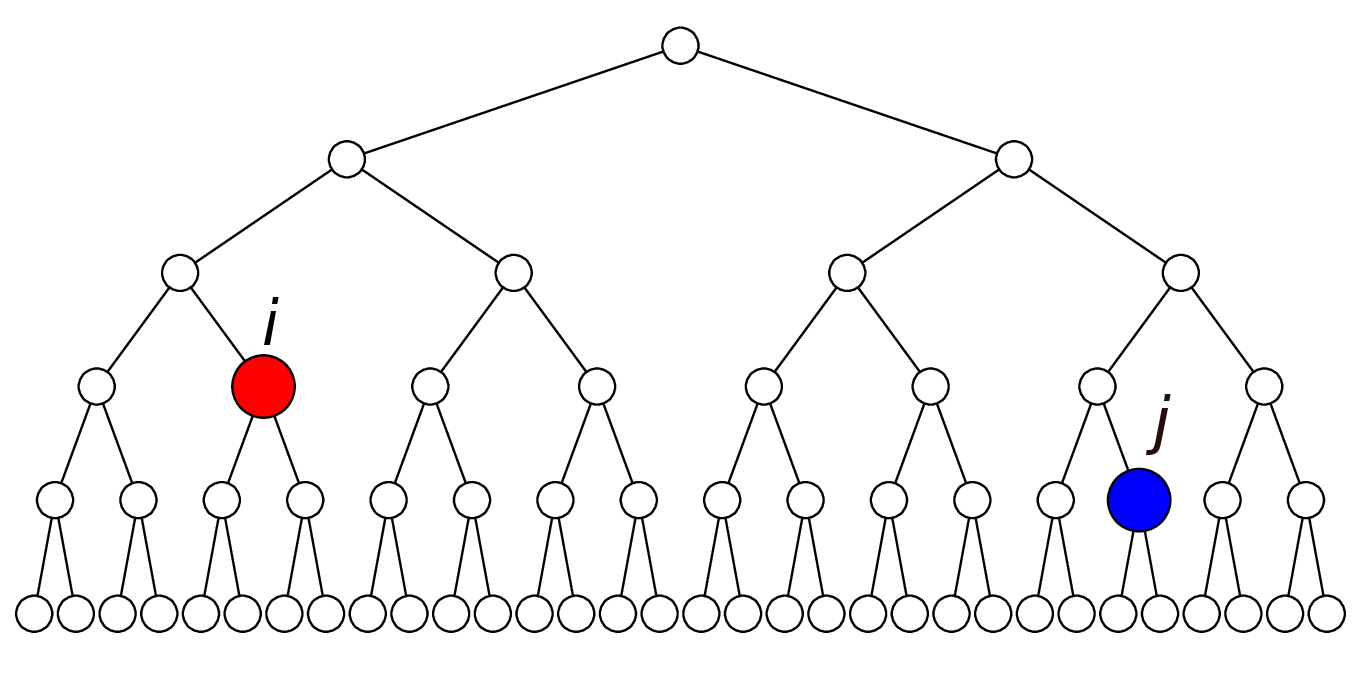}
    \caption{\label{fig_MI} Mutual information $I$ defined by Eq.~(\ref{eq_def_MI}) is the mutual information between the red node $i$ and the blue node $j$.
    }
\end{figure}

As described in the preceding section, we examined the distribution of a symbol on a node, but we could find no significant difference between a PCFG and a PCSG.
For the discussion in this section, we turn our interest to mutual information, which has a close relation to the two-point correlation function \cite{Li1990} and which has been used for measuring correlation in symbolic sequences such as formal and natural languages \cite{Li1989language, Lin2017, Tanaka2021}, music \cite{Lin2017}, birdsong \cite{Sainburg2019}, DNA \cite{Li1992}, and so forth.
We denote the mutual information between nodes $i$ and $j$, as depicted in Fig.~\ref{fig_MI}, as
\begin{equation}
    I_{i, j} ( q, M )
    \equiv \sum_{\sigma_i, \sigma_j} P \left( \sigma_i, \sigma_j \right)
    \ln \frac{
        P \left( \sigma_i, \sigma_j \right)
    }{
        P \left( \sigma_i \right) P \left( \sigma_j \right)
    } .
    \label{eq_def_MI}
\end{equation}
This measures the dependence between the two nodes.
Although the behavior of mutual information in a PCFG is well known through theoretical analysis \cite{Lin2017}, this analysis is also based on context-free independence.
Consequently, understanding what occurs in a PCSG regarding the mutual information, where the independence is broken, is non-trivial again.

\begin{figure}
    \centering
    \includegraphics[width = 1 \linewidth]{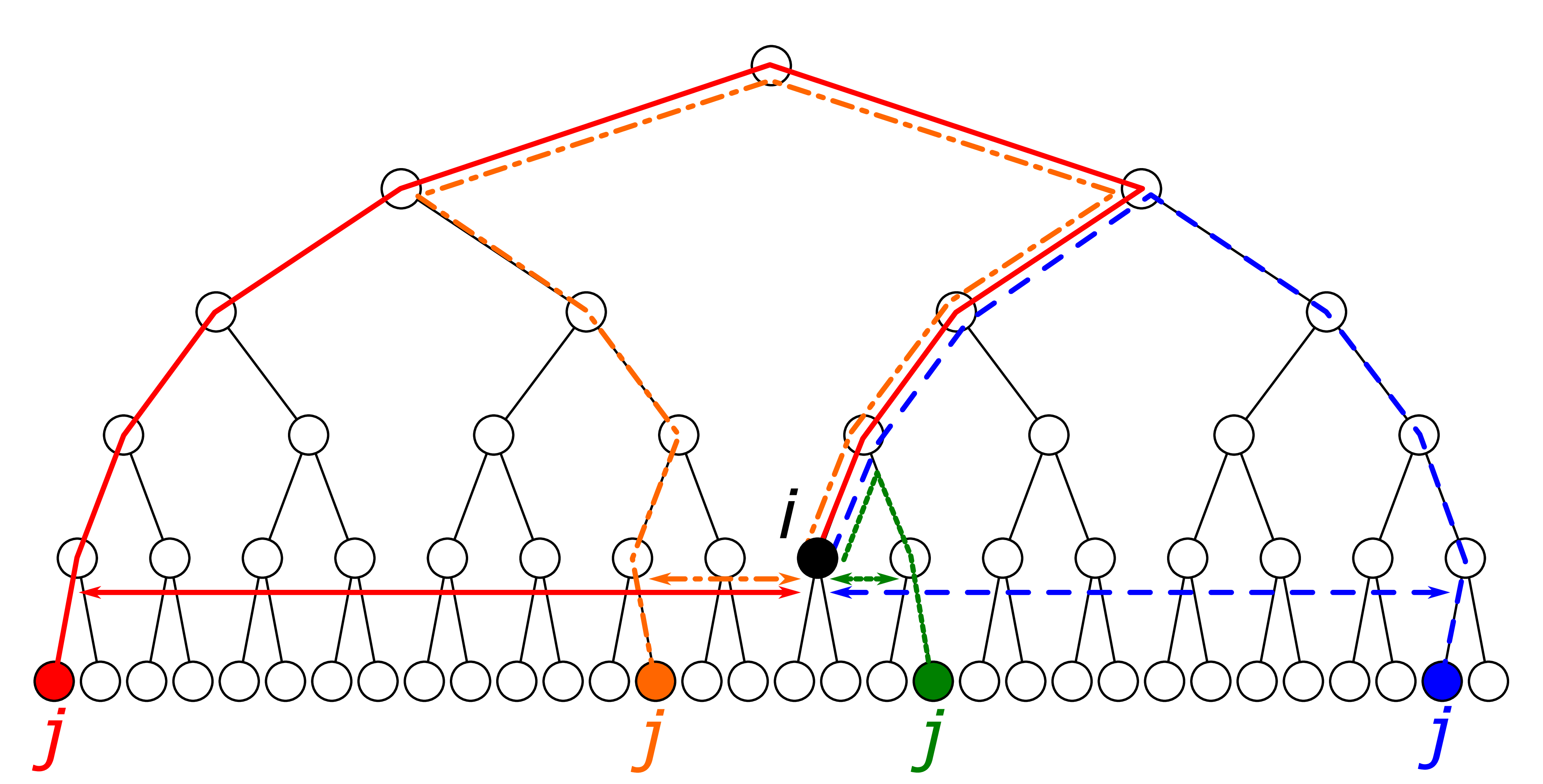}
    \caption{\label{fig_distances} Structural and horizontal distances between node $i$ and $j$ with $i = (1, 0, 0, 0)$ and $j = (0, 0, 0, 0, 0), (0, 1, 1, 0, 1), (1, 0, 0, 1, 1)$, or $(1, 1, 1, 1, 0)$.
        Different colors and lines represent different $j$s.
        The structural distance is the path length from $i$ to $j$, denoted by the line along the edges.
        The horizontal distance is the number of nodes lying horizontally between the higher node and the lower node's ancestor of the same depth as the former, as indicated by the horizontal arrows.
        Nodes $j = (0, 0, 0, 0, 0)$ and $ (0, 1, 1, 0, 1)$ are in the left subtree.
        The horizontal distance is 8 in the former case and 2 in the latter case, whereas the structural distance is 9 in both cases.
        Node $j = (1, 0, 0, 1, 1)$ belongs to the right subtree.
        The structural and horizontal distances between this node and $i$ are, respectively, 3 and 1.
        Node $j = (1, 1, 1, 1, 0)$ belongs to the right subtree, too.
        The structural and horizontal distances are 7.
        When $j$ belongs to the right branch, the horizontal distance grows exponentially as the structural distance increases.}
\end{figure}

Before presenting the results of the numerical analysis, we introduce some notations and quantities.
In the following, we designate a node by a binary sequence that represents the path from the root to the node by assigning 0 and 1, respectively, to a left and right child.
For example, nodes $()$, $(0)$, and $(0, 1)$ represent the root, the left child of the root, and the right child of the left child of the root, respectively.
To characterize the relative position of two nodes, we introduce the two distinct distances described in Fig.~\ref{fig_distances}.
The first is the structural distance, i.e., the length of the path between the two nodes.
The second, designated as the horizontal distance, is the number of nodes lying horizontally between the two nodes.
If the depths of the two nodes differ, then the horizontal distance is the number of nodes between the higher node and the lower node's ancestor of the same depth as the former.

One of the two nodes was fixed at $i = (1, 0, 0, 0, 0, 0)$, which is the leftmost node of depth $6$
in the subtree whose root is the right child of the root of the whole tree.
The other node $j$ could be any node in the whole tree.
The relation between structural and horizontal distances differs based on whether node $j$ belongs to the left or right subtree, as presented in Fig.~\ref{fig_distances}.
Presuming that the depth of node $j$ is fixed,
then when $j$ is in the left subtree, i.e., $j = (0, \cdots)$, the structural distance is the same, irrespective of the horizontal distance.
However, when $j$ is in the right subtree, i.e., $j = (1, \cdots)$, the horizontal distance is roughly exponential of the structural distance.

In the context-free case with $q = 0$, the dependence of the mutual information on the two distances is already known.
Lin and Tegmark \cite{Lin2017} have proved that the mutual information decays exponentially with the structural distance.
Recalling that the mutual information in a Markov chain decays exponentially with the chain length, this result is intuitively reasonable when considering context-free independence.
When $j$ is in the left subtree, the mutual information is the same for any node $j$ of the same depth because the mutual information depends only on the structural distance, which is independent of the horizontal distance.
However, when $j$ is in the right subtree, the mutual information decays according to a power law of the horizontal distance because the horizontal distance grows exponentially in the structural distance.
One main claim of Lin and Tegmark \cite{Lin2017} was that this power law might be the mechanism of the power-law decay of mutual information in natural language 
texts.

In the context-sensitive case with $q > 0$, we examined the behavior of the mutual information. We sampled $10^8$ complete trees of depth $D = 7$ and estimated $I$.
Because the mutual information between $X$ and $Y$ is decomposed into $S(X) + S(Y) - S(X, Y)$ where $S(\cdot)$ is Shannon entropy, we computed the mutual information by estimating the entropy from the empirical distribution. This estimate has a bias from the entropy of the true distribution, resulting in biased mutual information, which is not negligible in the region of the small mutual information. 
Consequently, to compute the entropy in the present and the subsequent sections, we used the bias-reduced estimator proposed by Ref.~\cite{Grassberger2003}. This estimator is represented by
\begin{equation*}
    \hat{S} (X)
    \equiv \Psi ( N ) - \frac{1}{N} \sum_x n_x \Psi ( n_x ).
\end{equation*}
Therein, $\Psi$ is the digamma function, $x$ represents a state which $X$ takes, $n_x$ denotes the number of samples such that $X = x$, and $N = \sum_x n_x$ is the total number of samples.

\begin{figure*}
    \centering
    \includegraphics[width = .7 \linewidth]{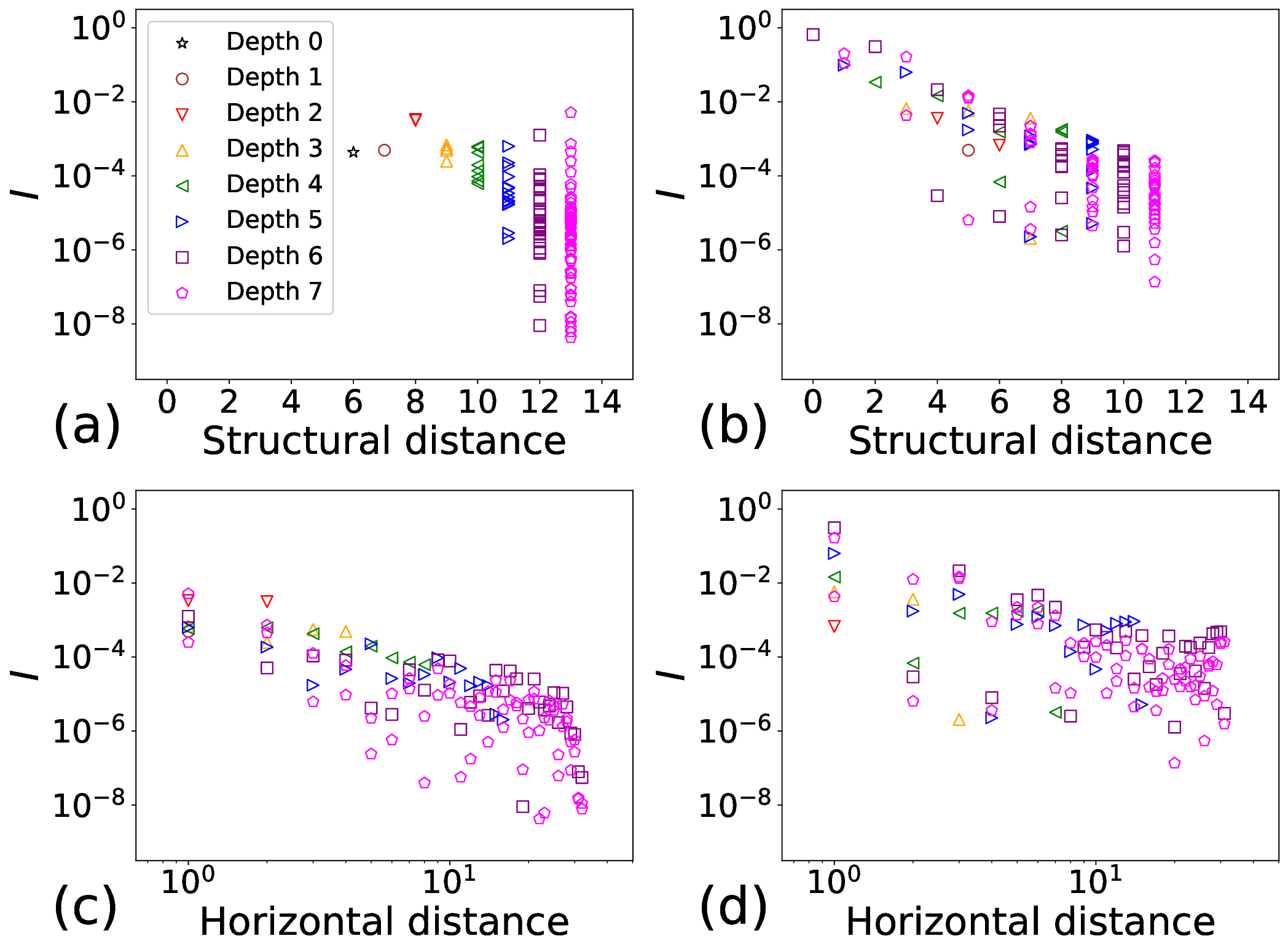}
    \caption{\label{fig_MI_individual} Mutual information $I$ defined by Eq.~(\ref{eq_def_MI}) against the distance between $i$ and $j$. Weights $M$ are generated with $\epsilon = 10^{-2}$. The context sensitivity is set to $q = 1$. The position of $i$ is fixed at $(1, 0, 0, 0, 0, 0)$. Mutual information against the structural distance when node $j$ is in the left branch, i.e., $j= (0, \cdots )$, is shown in (a). The same quantity when node $j$ is in the right branch, i.e., $j= (1, \cdots )$, is in (b). Similarly, plots against the horizontal distance are shown in (c) for $j = (0, \cdots)$ and (d) for $j = (1, \cdots)$. The result when node $j$ is the root, i.e., $j = ()$, is in (a). Markers and colors are different for different depths.
    }
\end{figure*}

\begin{figure}
    \centering
    \includegraphics[width = 1 \linewidth]{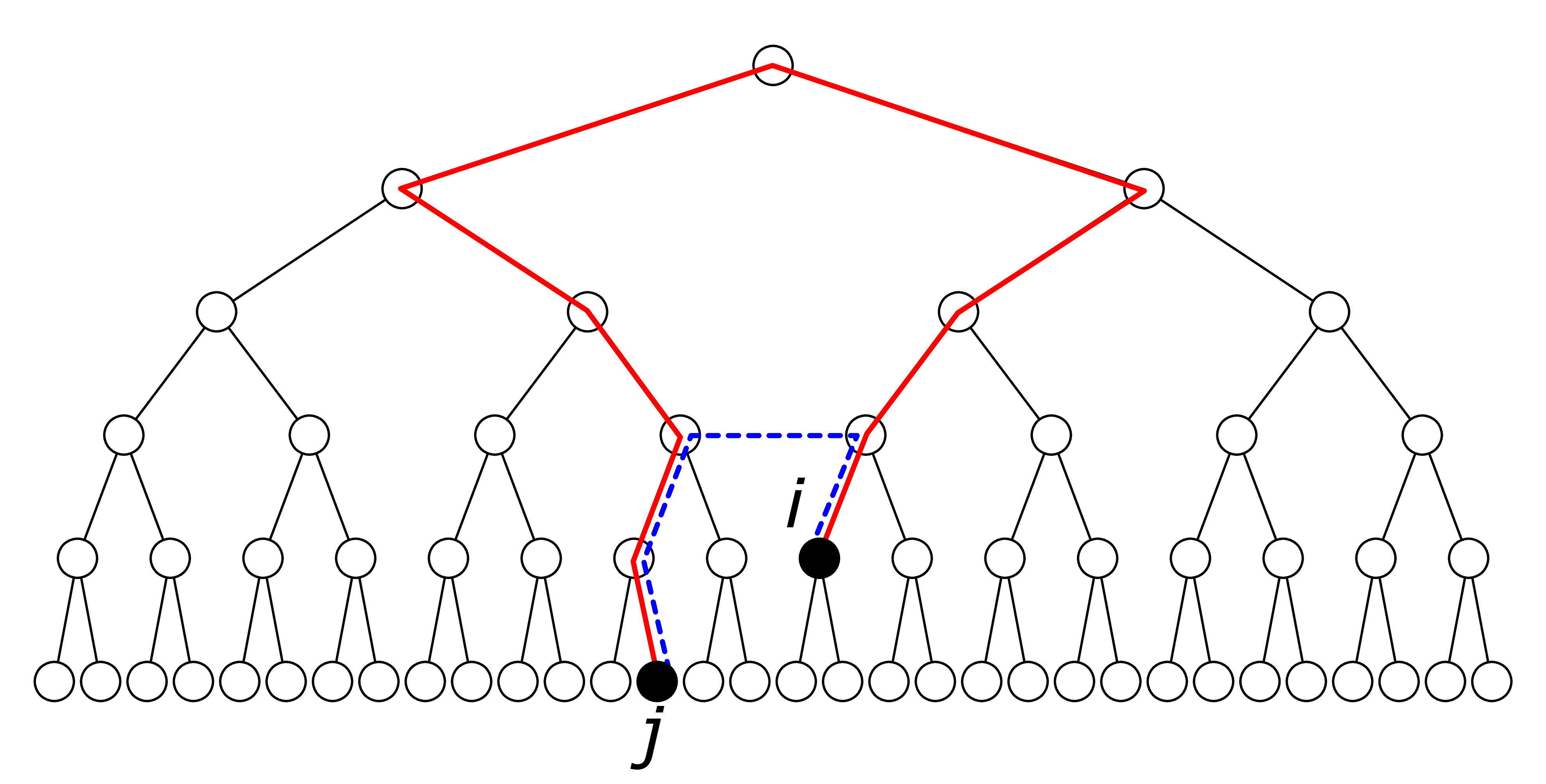}
    \caption{\label{fig_effective_distance}
        Structural and effective distances between nodes $i = (1, 0, 0, 0)$ and $j = (0, 1, 1, 0, 1)$.
        The former is described by the red line whereas the latter is shown by the blue dashed line.}
\end{figure}

Figure~\ref{fig_MI_individual} shows $I$s for an $M$ generated with $\epsilon = 10^{-2}$ and $q = 1$, where rewriting always refers to the context. The structural distance dependences of $I$ for $j$ belonging to the left and right branches are shown respectively in Figs.~\ref{fig_MI_individual}(a) and ~\ref{fig_MI_individual}(b). The horizontal distance dependence is also shown in Figs.~\ref{fig_MI_individual}(c) and ~\ref{fig_MI_individual}(d). 
When $j$ belongs to the 
right
branch, i.e., $j = (1, \cdots)$, as shown in the right subfigures (b) and (d), what is observed with a PCFG roughly holds. Here, $I$ decays exponentially in the structural distance and follows a power law in the horizontal distance. However, different behavior is observed when $j$ belongs to the left branch, i.e., $j = (0, \cdots)$, as shown in the left subfigures (a) and (c). In Fig.~\ref{fig_MI_individual}(a), $I$ has clearly different values even with the same structural distances, whereas it decays in the power law of the horizontal distance in Fig.~\ref{fig_MI_individual}(c), similarly to the case with $j = (1, \cdots)$. This result differs from the behavior found with a PCFG.

The mutual information between nodes in a PCSG depends explicitly on the horizontal distance.
This observation can be attributed to the context sensitivity inherent in PCSG rules.
If the context-free independence holds, then a node can correlate with other nodes only along the path in the tree graph.
This result engenders the exponential decay with the structural distance.
In contrast, in a PCSG where each rule involves the context $L$ and $R$ as well as $A$, a node can correlate with its left and right neighbors directly, even in the absence of a direct path between them.
This horizontal correlation can bypass the long structural distance between two nodes belonging to different subtrees, leading to the effective distance. As shown in Fig.~\ref{fig_effective_distance}, the horizontal distance increases exponentially with the effective distance.
If the mutual information does not decay exponentially with the structural distance, but instead with the effective distance, then the mutual information will decay in a power law in the horizontal distance, irrespective of whether node $j$ belongs to the left or the right branch.

\begin{figure*}
    \centering
    \includegraphics[width = 1 \linewidth]{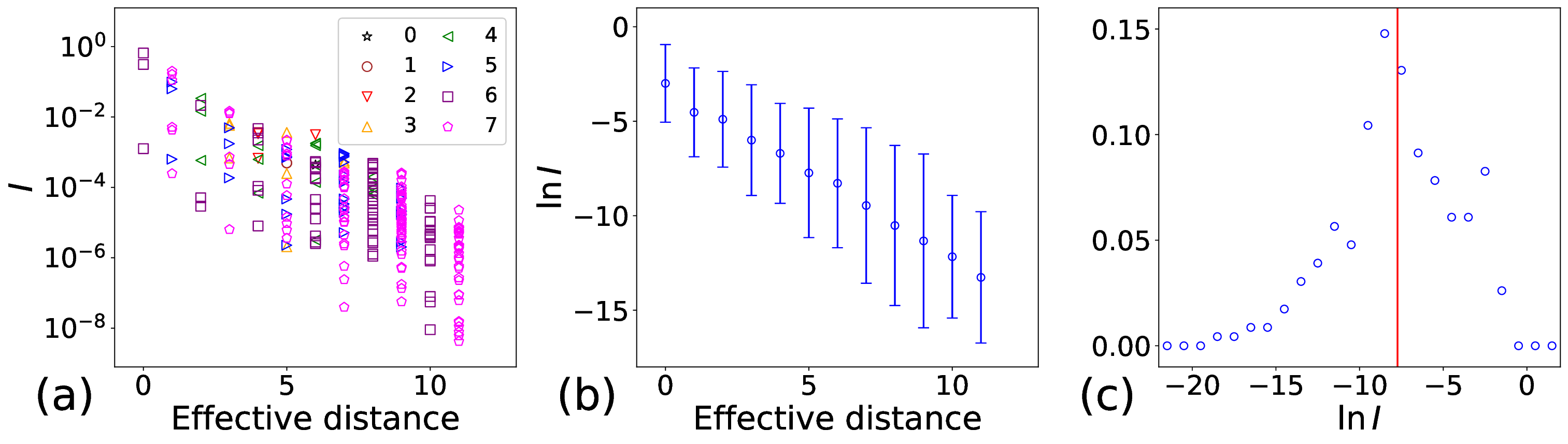}
    \caption{\label{fig_MI_effective}(a) Mutual information $I$ defined by Eq.~(\ref{eq_def_MI}) against the effective distance between $i$ and $j$.
        Weights $M$ are generated from the lognormal distribution. Markers and colors differ for different depths. (b) Averages and standard deviations of $\ln I$ over $j$s of the same effective distance and over 10 $M$s generated. (c) Normalized histograms of $\ln I$ for 10 $M$s for which the effective distance is 5. The red vertical line represents the average. For all (a), (b), and (c), the parameter in the lognormal distribution is $\epsilon = 10^{-2}$, the context sensitivity is $q = 1$, and node $i$ is fixed at $(1, 0, 0, 0, 0, 0)$.}
\end{figure*}

The effective distance is definable as explained hereinafter.
Presuming that nodes $i'$ and $j'$ are ancestors of $i$ and $j$, respectively, and that $i'$ and $j'$ are the horizontal neighbors of one another, then the effective distance between nodes $i$ and $j$ is the sum of the path length from $i$ to $i'$ and from $j$ to $j'$.
Here, we assume that the effective distance is equal to the structural distance if one of the two nodes is the ancestor of the other.
We plot the same $I$ as in Fig.~\ref{fig_MI_individual}, but against the effective distance, in Fig.~\ref{fig_MI_effective}(a).
From this, it can be confirmed that the mutual information decays exponentially with the effective distance, as expected.
This result indicates the existence of a typical effective distance that corresponds to a correlation length, which is the inverse of the decay rate. The mutual information is small beyond this typical distance.

It is intuitively reasonable to infer that the mutual information decays exponentially with the effective distance. Joint probability $P ( \sigma_0, \cdots, \sigma_{2^{D+1}-2} )$ of all nodes is the product of all $2^{D}-1$ rewriting weights. For two nodes $i$ and $j$, marginalizing the remaining nodes yields the joint probability $P ( \sigma_i, \sigma_j )$ of the two nodes. The greatest contribution to this is the product of the weights on the shortest effective path, described by the blue dashed line in Fig.~\ref{fig_effective_distance}. Although an effective path and its corresponding weights depend on the order of application of rules at each step, the length of the shortest effective path asymptotically equals the effective distance. Therefore, the joint probability of two nodes scales as an exponential function of the effective distance.
This result implies that the mutual information scales in the same manner \cite{Lin2017}.

What we describe here is not unique to this instance. It is typically observed across the $M$s sampled.
We measured $I$ for 10 $M$s under the same settings and computed the averages and the standard deviations of $\ln I$ over $j$s of each effective distance and over $M$s.
Whereas mutual information is always non-negative, the estimate by the method in \cite{Grassberger2003} sometimes takes negative values when the true value is small.
We simply excluded non-positive estimates to compute the logarithm.
This exclusion caused the average to be biased upward, but this bias was slight in this case.
The results presented in Fig.~\ref{fig_MI_effective}(b) show that the exponential decay in the effective distance discussed above for a single $M$ is observed across $10$ $M$s.
Figure~\ref{fig_MI_effective}(c) also presents histograms of $\ln I$ for the effective distance $5$, where the frequencies are normalized.
The points are distributed around the average.
The deviations in Fig.~\ref{fig_MI_effective}(b) and the distribution in Fig.~\ref{fig_MI_effective}(c) originate from differences in $j$s and $M$s rather than from sample fluctuations.

The rate of decay and the correlation length depend on weights $M$, causing the average rate to change as the parameter $\epsilon$ varies. One can infer that, as $\epsilon$ increases, the distribution of trees under generated weights $M$ tends to approach the uniform distribution. Therefore, the mutual information is expected to decay faster, meaning that the correlation length will become smaller. Additionally, the rate of decay depends on the context sensitivity $q$. With larger $q$, rewriting operations depend not only on the rewritten symbol but also on the context, with higher probability.
This dependence seems to engender slower decay. The numerically computed mutual information with different $\epsilon$ and $q$, as presented in Supplemental Material, 
follows these expectations.

\begin{figure}[t]
    \centering
    \includegraphics[width = 1 \linewidth]{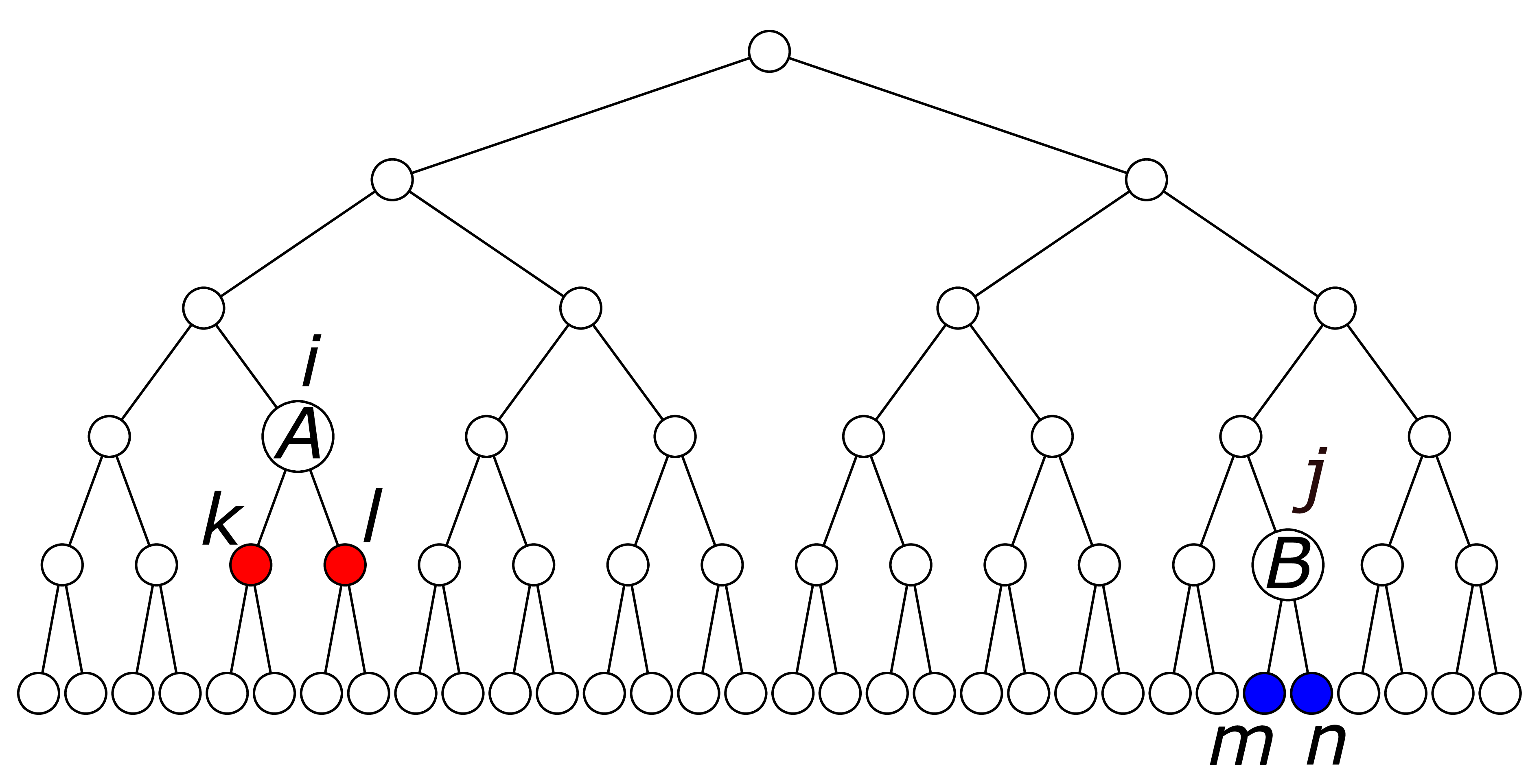}
    \caption{Context-free independence breaking \label{fig_CMI} $J$ defined by Eq.~(\ref{eq_def_CMI}) is the mutual information between the red nodes $k$ and $l$ and the blue nodes $m$ and $n$.
    }
\end{figure}

\section{Quantification of Context-free Independence Breaking}
\label{sec_violation}

\begin{figure*}
    \centering
    \includegraphics[width = 1 \linewidth]{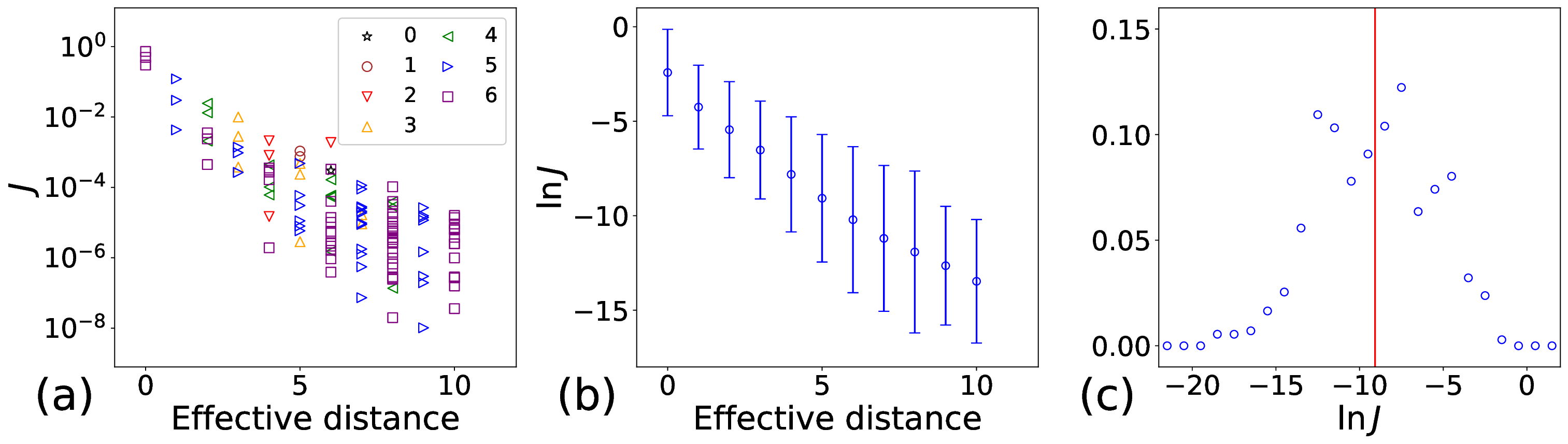}
    \caption{\label{fig_CMI_effective}(a) Degree of the context-free independence breaking, or parent-fixed mutual information $J$ defined by Eq.~(\ref{eq_def_CMI}) for $A = B = 0$ against the effective distance between $i$ and $j$.
        Weights $M$ are generated from the lognormal distribution.
        Markers and colors differ for different depths.
        (b) Averages and standard deviations of $\ln J$ over $j$s of the same effective distance, over the symbols $A$ of node $i$ and $B$ of $j$, and over 10 $M$s generated.
        (c) Normalized histograms of $\ln J$ for 10 $M$s with effective distance $5$.
        The red vertical line represents the average.
        For all (a), (b), and (c), the parameter in the lognormal distribution is $\epsilon = 10^{-2}$, the context sensitivity is $q = 1$, and node $i$ is fixed at $(1, 0, 0, 0, 0, 0)$.}
\end{figure*}

Finally, we investigate the effect of context sensitivity more directly by quantifying the extent to which the context-free independence is broken. This independence means that two subtrees are mutually independent under the condition that the symbols of their roots are fixed. Therefore, quantifying the breakage of the context-free independence involves the measurement of the mutual information between the subtrees under this condition. However, it requires extremely large amounts of data to obtain the distribution of a subtree when the subtree is large. To overcome this difficulty, we instead specifically examine the mutual information between the children of their roots, as shown in Fig.~\ref{fig_CMI}.
We denote this mutual information as
\begin{widetext}
    \begin{equation}
        J_{i, j; A, B} ( q, M )
        \equiv \sum_{
            \sigma_k, \sigma_l, \sigma_m, \sigma_n
        } P \left(
            \sigma_k, \sigma_l, \sigma_m, \sigma_n |
            \sigma_i = A, \sigma_j = B
        \right)
        \ln \frac{
            P \left(
                \sigma_k, \sigma_l, \sigma_m, \sigma_n |
                \sigma_i = A, \sigma_j = B
            \right)
        }{
            P \left(
                \sigma_k, \sigma_l |
                \sigma_i = A, \sigma_j = B
            \right)
            P \left(
                \sigma_m, \sigma_n |
                \sigma_i = A, \sigma_j = B
            \right)
        },
        \label{eq_def_CMI}
    \end{equation}
\end{widetext}
Therein, $k$ and $l$ respectively represent the left and the right children of $i$; $m$ and $n$ are the children of $j$.
This quantity is always zero for any $i$ and $j$ in a PCFG because of the context-free independence. This is the requirement that must be met for this quantity to be a meaningful metric of the breaking of independence.

In addition to measuring the degree of context-free independence breaking, the metric $J$ has other interpretations. One interpretation derives from theoretical physics. If the network of interactions forms a tree, where every interaction in the system is between a node and its child, then $J$ is zero. In the presence of loops in the network, as seen in a PCSG, $J$ can take a positive value. In this sense, $J$ represents the degree to which the network of interactions deviates from a tree. Another interpretation is linguistic: Suppose that two constituents or phrases, i.e. subtrees of a derivation, are, for example, a noun phrase and a verb phrase.
Under this condition, the structures of the noun phrase and the verb phrase are mutually dependent; $J$ represents the strength of this dependence.

We measured the context-free independence breaking $J$ in the same manner as for the mutual information $I$ in the preceding section, under the same setting, where $i = (1, 0, 0, 0, 0, 0)$, $q = 1$, $\epsilon = 10^{-2}$, and $10^8$ trees of the depth $7$ were sampled for each $M$.
Our observations revealed that 
$J$ behaves very similarly to $I$.
Figure~\ref{fig_CMI_effective}(a) shows $J$ for $A = B = 0$ against the effective distance for an $M$ generated from the lognormal distribution.
It is evident that $J$ exhibits exponential decay with the effective distance.
Again, this finding indicates that there exists a correlation length, or a typical effective distance beyond which the dependence between two subtrees is small.
We computed the averages and the standard deviations of $\ln J$ over $j$'s of each effective distance, over $M$'s, and over $A$ and $B$, using the data size, i.e., the number of generated trees satisfying $\sigma_i = A$ and $\sigma_j = B$, as the weights.
Additionally, we simply discarded non-positive estimates of $J$,
which only led to a small bias.
Figure~\ref{fig_CMI_effective}(b) presents the results, suggesting that the exponential decay of $J$ with the effective distance occurs across different $M$s, as well as different $A$s and $B$s.
Figure~\ref{fig_CMI_effective}(c) shows the normalized histogram of $\ln J$ obtained for the effective distance $5$, where the data sizes were used as the weights. The distribution of $\ln J$s centers around the red vertical line representing the average.

The dependence of $J$ on the parameter $\epsilon$ and the context sensitivity $q$ exhibits similar tendencies to those observed for $I$. Particularly, as $\epsilon$ increases or $q$ decreases, the decay rate becomes more pronounced whereas the correlation length becomes smaller. Supplemental Materials 
provide additional results for different values of $\epsilon$ and $q$.
In a general system, $I$ and $J$ do not necessarily behave similarly. Indeed, in a PCFG, $I$ is positive and decays exponentially with the structural distance, whereas $J$ is always zero. It is somewhat non-trivial that both $I$ and $J$ decay exponentially with the effective distance in a PCSG.

\section{Conclusion}
\label{sec_conclusion}

A PCFG, a simple mathematical model for randomly generating a tree, has been used to model various hierarchical phenomena, including natural languages. This model satisfies the assumptions of context-free independence. Although this feature allows for the theoretical analysis of various properties of a PCFG, the restriction is too strong for a PCFG to be expressive of distributions.

We introduced the simple PCSG by relaxing the context-free independence, and we analyzed its statistical properties systematically. First, we specifically examined the distribution of a symbol on a single node. This distribution is to a PCSG what magnetization is to a spin system. Although the context sensitivity affects the distribution, its effect brings only continuous and quantitative changes. Such changes can occur even without context sensitivity, for example in the interpolation between two PCFGs.
Our numerical investigation also shows that the Binder parameter of the mean ratio of a symbol, which is an analytic function of $\epsilon$ in the context-free RLM, is unlikely to be discontinuous in a context-sensitive case.

The second quantity of interest is the mutual information between two nodes, which is related closely to the two-point correlation function \cite{Li1990}.
It is noteworthy that mutual information decays exponentially with the effective distance between two nodes,
which is a consequence of the horizontal correlation because of context sensitivity.
This feature contrasts with the fact that the decay of the mutual information in a PCFG is exponential with respect to the structural distance, i.e., the path length.

In addition, to quantify the degree to which context-free independence is broken, we proposed the use of mutual information between two pairs of nodes under the condition that the parent symbols are fixed. This metric can also indicate the degree to which the network of interactions deviates from a tree in theoretical physics, and it can indicate the mutual dependence between the structures of two constituents in linguistics. This quantity emphasizes the most distinct difference between a PCFG and a PCSG. The context-free independence breaking decays exponentially with the effective distance in a PCSG, similar to the mutual information between two nodes, whereas the breaking always remains zero in a PCFG.

Possible future issues, in our view, are divisible into four main directions. First, it is necessary to develop methods for theoretical analysis and efficient numerical approximation to confirm and further investigate the behaviors of PCSGs observed in this study, such as the exponential decay of the mutual information and the context-free independence breaking. The main challenges are the exponential growth of tree sizes and the complex interactions due to context sensitivity.

Second, another important approach would be to examine specific PCSGs, particularly those exhibiting atypical behavior, in contrast to our analysis of the typical properties of randomly sampled PCSGs.
It might be true that PCSGs with low probabilistic measures exhibit non-analytic behavior in $\Delta$ as a function of the context sensitivity $q$, or non-exponential decay of the mutual information or context-free independence breaking. The existence of such PCSGs and the mechanism underlying their atypical behavior are left as intriguing open problems.

Third, CSG is not the only linguistic framework beyond CFG. Although the CSG framework makes tree structures context-sensitive in a straightforward manner, modern linguists do not consider a CSG to be a relevant model of a natural language. This skepticism arises because a CSG can generate a set of sentences extending beyond natural languages \cite{jager2012formal}. Also, formal language theory predominantly addresses surface sentences rather than syntactic structures \cite{Fukui2015}. Conversely, several alternative models have been proposed as grammars closer to natural languages, such as Tree Adjoining Grammar \cite{Joshi1985}, Combinatory Categorial Grammar \cite{Steedman1987}, and Minimalist Grammar \cite{Stabler1997}. The natural progression is to introduce probabilistic extensions to these grammars and to investigate their statistical properties, as examined in this study. Particularly, all probabilistic extensions of a CSG and the three grammars described above will violate the context-free independence, but their independence breaking $J$ might decay exponentially, polynomial, or non-monotonically, depending on the grammar. If the decay is, for example, exponential in every model, then their decay rates might differ. These probabilistic grammars can be characterized by emphasizing the distinctions in their independence breaking $J$, thereby contributing to a comprehensive understanding of the grammars from a physical perspective.

As a fourth point, we discuss the application of our metric $J$ for the context-free independence breaking, which is applicable not only to probabilistic grammars such as PCFGs but also to any distribution of a tree, including those underlying human languages and birdsongs. Earlier research has demonstrated that the behavior of mutual information in PCFGs, human languages, and birdsongs is similar in that it decays as a power-law function of the horizontal distance or the sequence length \cite{Lin2017, Sainburg2019}. 
However, $J$ will allow us to detect and quantify the distinction between human languages and PCFGs, given the empirical knowledge that context-free independence breaking occurs in natural languages \cite{Johnson2006, ODonnell2009}. It might also be possible to identify characteristics unique to human languages, which are not present in birdsongs, using $J$. By quantifying the degree of independence breaking, we can more deeply compare tree structures among different mathematical models or natural phenomena.

\begin{acknowledgments}
    We would like to thank R. Yoshida, K. Kajikawa, Y. Oseki, Y. Toji, J. Takahashi, and H. Miyahara for useful discussions. This work was supported by JSPS KAKENHI Grant Nos. 23KJ0622 and 23H01095, JST Grant No. JPMJPF2221, and the World-Leading Innovative Graduate Study Program for Advanced Basic Science Course at the University of Tokyo.
\end{acknowledgments}

\bibliography{2023_PCSG_PRR_published}

\providecommand{\noopsort}[1]{}\providecommand{\singleletter}[1]{#1}%
\begin{thebibliography}{47}%
\makeatletter
\providecommand \@ifxundefined [1]{%
 \@ifx{#1\undefined}
}%
\providecommand \@ifnum [1]{%
 \ifnum #1\expandafter \@firstoftwo
 \else \expandafter \@secondoftwo
 \fi
}%
\providecommand \@ifx [1]{%
 \ifx #1\expandafter \@firstoftwo
 \else \expandafter \@secondoftwo
 \fi
}%
\providecommand \natexlab [1]{#1}%
\providecommand \enquote  [1]{``#1''}%
\providecommand \bibnamefont  [1]{#1}%
\providecommand \bibfnamefont [1]{#1}%
\providecommand \citenamefont [1]{#1}%
\providecommand \href@noop [0]{\@secondoftwo}%
\providecommand \href [0]{\begingroup \@sanitize@url \@href}%
\providecommand \@href[1]{\@@startlink{#1}\@@href}%
\providecommand \@@href[1]{\endgroup#1\@@endlink}%
\providecommand \@sanitize@url [0]{\catcode `\\12\catcode `\$12\catcode `\&12\catcode `\#12\catcode `\^12\catcode `\_12\catcode `\%12\relax}%
\providecommand \@@startlink[1]{}%
\providecommand \@@endlink[0]{}%
\providecommand \url  [0]{\begingroup\@sanitize@url \@url }%
\providecommand \@url [1]{\endgroup\@href {#1}{\urlprefix }}%
\providecommand \urlprefix  [0]{URL }%
\providecommand \Eprint [0]{\href }%
\providecommand \doibase [0]{https://doi.org/}%
\providecommand \selectlanguage [0]{\@gobble}%
\providecommand \bibinfo  [0]{\@secondoftwo}%
\providecommand \bibfield  [0]{\@secondoftwo}%
\providecommand \translation [1]{[#1]}%
\providecommand \BibitemOpen [0]{}%
\providecommand \bibitemStop [0]{}%
\providecommand \bibitemNoStop [0]{.\EOS\space}%
\providecommand \EOS [0]{\spacefactor3000\relax}%
\providecommand \BibitemShut  [1]{\csname bibitem#1\endcsname}%
\let\auto@bib@innerbib\@empty
\bibitem [{\citenamefont {Chomsky}(1957)}]{Chomsky_SS}%
  \BibitemOpen
  \bibfield  {author} {\bibinfo {author} {\bibfnamefont {N.}~\bibnamefont {Chomsky}},\ }\href@noop {} {\emph {\bibinfo {title} {Syntactic Structures}}}\ (\bibinfo  {publisher} {Mouton \& Co.},\ \bibinfo {address} {Berlin},\ \bibinfo {year} {1957})\BibitemShut {NoStop}%
\bibitem [{\citenamefont {Jelinek}\ \emph {et~al.}(1992)\citenamefont {Jelinek}, \citenamefont {Lafferty},\ and\ \citenamefont {Mercer}}]{Jelinek1992}%
  \BibitemOpen
  \bibfield  {author} {\bibinfo {author} {\bibfnamefont {F.}~\bibnamefont {Jelinek}}, \bibinfo {author} {\bibfnamefont {J.~D.}\ \bibnamefont {Lafferty}},\ and\ \bibinfo {author} {\bibfnamefont {R.~L.}\ \bibnamefont {Mercer}},\ }\bibfield  {title} {\bibinfo {title} {Basic methods of probabilistic context free grammars},\ }in\ \href@noop {} {\emph {\bibinfo {booktitle} {Speech Recognition and Understanding}}},\ \bibinfo {editor} {edited by\ \bibinfo {editor} {\bibfnamefont {P.}~\bibnamefont {Laface}}\ and\ \bibinfo {editor} {\bibfnamefont {R.}~\bibnamefont {De~Mori}}}\ (\bibinfo  {publisher} {Springer Berlin Heidelberg},\ \bibinfo {address} {Berlin, Heidelberg},\ \bibinfo {year} {1992})\ pp.\ \bibinfo {pages} {345--360}\BibitemShut {NoStop}%
\bibitem [{\citenamefont {Charniak}(1997)}]{Charniak1997}%
  \BibitemOpen
  \bibfield  {author} {\bibinfo {author} {\bibfnamefont {E.}~\bibnamefont {Charniak}},\ }\bibfield  {title} {\bibinfo {title} {Statistical techniques for natural language parsing},\ }\href@noop {} {\bibfield  {journal} {\bibinfo  {journal} {AI {M}agazine}\ }\textbf {\bibinfo {volume} {18}},\ \bibinfo {pages} {33} (\bibinfo {year} {1997})}\BibitemShut {NoStop}%
\bibitem [{\citenamefont {Ellis}\ \emph {et~al.}(2015)\citenamefont {Ellis}, \citenamefont {Solar-Lezama},\ and\ \citenamefont {Tenenbaum}}]{Ellis2015}%
  \BibitemOpen
  \bibfield  {author} {\bibinfo {author} {\bibfnamefont {K.}~\bibnamefont {Ellis}}, \bibinfo {author} {\bibfnamefont {A.}~\bibnamefont {Solar-Lezama}},\ and\ \bibinfo {author} {\bibfnamefont {J.~B.}\ \bibnamefont {Tenenbaum}},\ }\bibfield  {title} {\bibinfo {title} {Unsupervised learning by program synthesis},\ }in\ \href@noop {} {\emph {\bibinfo {booktitle} {Adv.\ Neural Inf.\ Process.\ Syst.}}}\ (\bibinfo {year} {2015})\BibitemShut {NoStop}%
\bibitem [{\citenamefont {Worth}\ and\ \citenamefont {Stepney}(2005)}]{Worth2005}%
  \BibitemOpen
  \bibfield  {author} {\bibinfo {author} {\bibfnamefont {P.}~\bibnamefont {Worth}}\ and\ \bibinfo {author} {\bibfnamefont {S.}~\bibnamefont {Stepney}},\ }\bibfield  {title} {\bibinfo {title} {Growing music: Musical interpretations of l-systems},\ }in\ \href@noop {} {\emph {\bibinfo {booktitle} {Applications of Evolutionary Computing}}},\ \bibinfo {editor} {edited by\ \bibinfo {editor} {\bibfnamefont {F.}~\bibnamefont {Rothlauf}}, \bibinfo {editor} {\bibfnamefont {J.}~\bibnamefont {Branke}}, \bibinfo {editor} {\bibfnamefont {S.}~\bibnamefont {Cagnoni}}, \bibinfo {editor} {\bibfnamefont {D.~W.}\ \bibnamefont {Corne}}, \bibinfo {editor} {\bibfnamefont {R.}~\bibnamefont {Drechsler}}, \bibinfo {editor} {\bibfnamefont {Y.}~\bibnamefont {Jin}}, \bibinfo {editor} {\bibfnamefont {P.}~\bibnamefont {Machado}}, \bibinfo {editor} {\bibfnamefont {E.}~\bibnamefont {Marchiori}}, \bibinfo {editor} {\bibfnamefont {J.}~\bibnamefont {Romero}}, \bibinfo {editor} {\bibfnamefont {G.~D.}\ \bibnamefont {Smith}},\ and\ \bibinfo
  {editor} {\bibfnamefont {G.}~\bibnamefont {Squillero}}}\ (\bibinfo  {publisher} {Springer Berlin Heidelberg},\ \bibinfo {year} {2005})\ pp.\ \bibinfo {pages} {545--550}\BibitemShut {NoStop}%
\bibitem [{\citenamefont {Gilbert}\ and\ \citenamefont {Conklin}(2007)}]{Gilbert2007}%
  \BibitemOpen
  \bibfield  {author} {\bibinfo {author} {\bibfnamefont {{\'E}.}~\bibnamefont {Gilbert}}\ and\ \bibinfo {author} {\bibfnamefont {D.}~\bibnamefont {Conklin}},\ }\bibfield  {title} {\bibinfo {title} {A probabilistic context-free grammar for melodic reduction},\ }in\ \href@noop {} {\emph {\bibinfo {booktitle} {Proceedings of the International Workshop on Artificial Intelligence and Music, 20th International Joint Conference on Artificial Intelligence}}}\ (\bibinfo {address} {Hyderabad, India},\ \bibinfo {year} {2007})\ pp.\ \bibinfo {pages} {83--94}\BibitemShut {NoStop}%
\bibitem [{\citenamefont {Tano}\ \emph {et~al.}(2020)\citenamefont {Tano}, \citenamefont {Romano}, \citenamefont {Sigman}, \citenamefont {Salles},\ and\ \citenamefont {Figueira}}]{Tano2020}%
  \BibitemOpen
  \bibfield  {author} {\bibinfo {author} {\bibfnamefont {P.}~\bibnamefont {Tano}}, \bibinfo {author} {\bibfnamefont {S.}~\bibnamefont {Romano}}, \bibinfo {author} {\bibfnamefont {M.}~\bibnamefont {Sigman}}, \bibinfo {author} {\bibfnamefont {A.}~\bibnamefont {Salles}},\ and\ \bibinfo {author} {\bibfnamefont {S.}~\bibnamefont {Figueira}},\ }\bibfield  {title} {\bibinfo {title} {Towards a more flexible language of thought: Bayesian grammar updates after each concept exposure},\ }\href@noop {} {\bibfield  {journal} {\bibinfo  {journal} {Phys. Rev. E.}\ }\textbf {\bibinfo {volume} {101}},\ \bibinfo {pages} {042128} (\bibinfo {year} {2020})}\BibitemShut {NoStop}%
\bibitem [{\citenamefont {Lin}\ and\ \citenamefont {Tegmark}(2017)}]{Lin2017}%
  \BibitemOpen
  \bibfield  {author} {\bibinfo {author} {\bibfnamefont {H.~W.}\ \bibnamefont {Lin}}\ and\ \bibinfo {author} {\bibfnamefont {M.}~\bibnamefont {Tegmark}},\ }\bibfield  {title} {\bibinfo {title} {Critical behavior in physics and probabilistic formal languages},\ }\href@noop {} {\bibfield  {journal} {\bibinfo  {journal} {Entropy}\ }\textbf {\bibinfo {volume} {19}},\ \bibinfo {pages} {299} (\bibinfo {year} {2017})}\BibitemShut {NoStop}%
\bibitem [{\citenamefont {Knudsen}\ and\ \citenamefont {Hein}(1999)}]{Knudsen1999}%
  \BibitemOpen
  \bibfield  {author} {\bibinfo {author} {\bibfnamefont {B.}~\bibnamefont {Knudsen}}\ and\ \bibinfo {author} {\bibfnamefont {J.}~\bibnamefont {Hein}},\ }\bibfield  {title} {\bibinfo {title} {{RNA} secondary structure prediction using stochastic context-free grammars and evolutionary history},\ }\href@noop {} {\bibfield  {journal} {\bibinfo  {journal} {Bioinformatics}\ }\textbf {\bibinfo {volume} {15}},\ \bibinfo {pages} {446} (\bibinfo {year} {1999})}\BibitemShut {NoStop}%
\bibitem [{\citenamefont {Harlow}\ \emph {et~al.}(2012)\citenamefont {Harlow}, \citenamefont {Shenker}, \citenamefont {Stanford},\ and\ \citenamefont {Susskind}}]{Harlow2012}%
  \BibitemOpen
  \bibfield  {author} {\bibinfo {author} {\bibfnamefont {D.}~\bibnamefont {Harlow}}, \bibinfo {author} {\bibfnamefont {S.~H.}\ \bibnamefont {Shenker}}, \bibinfo {author} {\bibfnamefont {D.}~\bibnamefont {Stanford}},\ and\ \bibinfo {author} {\bibfnamefont {L.}~\bibnamefont {Susskind}},\ }\bibfield  {title} {\bibinfo {title} {Tree-like structure of eternal inflation: A solvable model},\ }\href@noop {} {\bibfield  {journal} {\bibinfo  {journal} {Phys. Rev. D}\ }\textbf {\bibinfo {volume} {85}},\ \bibinfo {pages} {063516} (\bibinfo {year} {2012})}\BibitemShut {NoStop}%
\bibitem [{\citenamefont {Li}(1989{\natexlab{a}})}]{Li1989open}%
  \BibitemOpen
  \bibfield  {author} {\bibinfo {author} {\bibfnamefont {W.}~\bibnamefont {Li}},\ }\bibfield  {title} {\bibinfo {title} {Spatial 1/f spectra in open dynamical systems},\ }\href@noop {} {\bibfield  {journal} {\bibinfo  {journal} {Europhys.\ Lett.}\ }\textbf {\bibinfo {volume} {10}},\ \bibinfo {pages} {395} (\bibinfo {year} {1989}{\natexlab{a}})}\BibitemShut {NoStop}%
\bibitem [{\citenamefont {Li}(1991)}]{Li1991}%
  \BibitemOpen
  \bibfield  {author} {\bibinfo {author} {\bibfnamefont {W.}~\bibnamefont {Li}},\ }\bibfield  {title} {\bibinfo {title} {Expansion-modification systems: A model for spatial 1/f spectra},\ }\href@noop {} {\bibfield  {journal} {\bibinfo  {journal} {Phys. Rev. A}\ }\textbf {\bibinfo {volume} {43}},\ \bibinfo {pages} {5240} (\bibinfo {year} {1991})}\BibitemShut {NoStop}%
\bibitem [{\citenamefont {Lieck}\ and\ \citenamefont {Rohrmeier}(2021)}]{Lieck2021}%
  \BibitemOpen
  \bibfield  {author} {\bibinfo {author} {\bibfnamefont {R.}~\bibnamefont {Lieck}}\ and\ \bibinfo {author} {\bibfnamefont {M.}~\bibnamefont {Rohrmeier}},\ }\bibfield  {title} {\bibinfo {title} {Recursive bayesian networks: Generalising and unifying probabilistic context-free grammars and dynamic bayesian networks},\ }in\ \href@noop {} {\emph {\bibinfo {booktitle} {Advances in Neural Information Processing Systems}}},\ Vol.~\bibinfo {volume} {34},\ \bibinfo {editor} {edited by\ \bibinfo {editor} {\bibfnamefont {M.}~\bibnamefont {Ranzato}}, \bibinfo {editor} {\bibfnamefont {A.}~\bibnamefont {Beygelzimer}}, \bibinfo {editor} {\bibfnamefont {Y.}~\bibnamefont {Dauphin}}, \bibinfo {editor} {\bibfnamefont {P.}~\bibnamefont {Liang}},\ and\ \bibinfo {editor} {\bibfnamefont {J.~W.}\ \bibnamefont {Vaughan}}}\ (\bibinfo  {publisher} {Curran Associates, Inc.},\ \bibinfo {year} {2021})\ pp.\ \bibinfo {pages} {4370--4383}\BibitemShut {NoStop}%
\bibitem [{\citenamefont {Lindenmayer}(1968{\natexlab{a}})}]{Lindenmayer1968I}%
  \BibitemOpen
  \bibfield  {author} {\bibinfo {author} {\bibfnamefont {A.}~\bibnamefont {Lindenmayer}},\ }\bibfield  {title} {\bibinfo {title} {Mathematical models for cellular interactions in development i. filaments with one-sided inputs},\ }\href@noop {} {\bibfield  {journal} {\bibinfo  {journal} {Journal of Theoretical Biology}\ }\textbf {\bibinfo {volume} {18}},\ \bibinfo {pages} {280} (\bibinfo {year} {1968}{\natexlab{a}})}\BibitemShut {NoStop}%
\bibitem [{\citenamefont {Lindenmayer}(1968{\natexlab{b}})}]{Lindenmayer1968II}%
  \BibitemOpen
  \bibfield  {author} {\bibinfo {author} {\bibfnamefont {A.}~\bibnamefont {Lindenmayer}},\ }\bibfield  {title} {\bibinfo {title} {Mathematical models for cellular interactions in development ii. simple and branching filaments with two-sided inputs},\ }\href@noop {} {\bibfield  {journal} {\bibinfo  {journal} {Journal of Theoretical Biology}\ }\textbf {\bibinfo {volume} {18}},\ \bibinfo {pages} {300} (\bibinfo {year} {1968}{\natexlab{b}})}\BibitemShut {NoStop}%
\bibitem [{\citenamefont {Herman}\ and\ \citenamefont {Walker}(1974)}]{Herman1974}%
  \BibitemOpen
  \bibfield  {author} {\bibinfo {author} {\bibfnamefont {G.~T.}\ \bibnamefont {Herman}}\ and\ \bibinfo {author} {\bibfnamefont {A.}~\bibnamefont {Walker}},\ }\bibfield  {title} {\bibinfo {title} {Context free languages in biological systems},\ }\href@noop {} {\bibfield  {journal} {\bibinfo  {journal} {International Journal of Computer Mathematics}\ }\textbf {\bibinfo {volume} {4}},\ \bibinfo {pages} {369} (\bibinfo {year} {1974})}\BibitemShut {NoStop}%
\bibitem [{\citenamefont {Nakaishi}\ and\ \citenamefont {Hukushima}(2022)}]{Nakaishi22}%
  \BibitemOpen
  \bibfield  {author} {\bibinfo {author} {\bibfnamefont {K.}~\bibnamefont {Nakaishi}}\ and\ \bibinfo {author} {\bibfnamefont {K.}~\bibnamefont {Hukushima}},\ }\bibfield  {title} {\bibinfo {title} {Absence of phase transition in random language model},\ }\href@noop {} {\bibfield  {journal} {\bibinfo  {journal} {Phys.\ Rev.\ Reas.}\ }\textbf {\bibinfo {volume} {4}},\ \bibinfo {pages} {023156} (\bibinfo {year} {2022})}\BibitemShut {NoStop}%
\bibitem [{\citenamefont {Booth}\ and\ \citenamefont {Thompson}(1973)}]{Booth1973}%
  \BibitemOpen
  \bibfield  {author} {\bibinfo {author} {\bibfnamefont {T.~L.}\ \bibnamefont {Booth}}\ and\ \bibinfo {author} {\bibfnamefont {R.~A.}\ \bibnamefont {Thompson}},\ }\bibfield  {title} {\bibinfo {title} {Applying probability measures to abstract languages},\ }\href@noop {} {\bibfield  {journal} {\bibinfo  {journal} {IEEE Trans. Comput.}\ }\textbf {\bibinfo {volume} {C-22}},\ \bibinfo {pages} {442} (\bibinfo {year} {1973})}\BibitemShut {NoStop}%
\bibitem [{\citenamefont {Miller}\ and\ \citenamefont {O'Sullivan}(1992)}]{Miller1992}%
  \BibitemOpen
  \bibfield  {author} {\bibinfo {author} {\bibfnamefont {M.~I.}\ \bibnamefont {Miller}}\ and\ \bibinfo {author} {\bibfnamefont {J.~A.}\ \bibnamefont {O'Sullivan}},\ }\bibfield  {title} {\bibinfo {title} {Entropies and combinatorics of random branching processes and context-free languages},\ }\href@noop {} {\bibfield  {journal} {\bibinfo  {journal} {IEEE Trans.\ Inf.\ Theor.}\ }\textbf {\bibinfo {volume} {38}},\ \bibinfo {pages} {1292} (\bibinfo {year} {1992})}\BibitemShut {NoStop}%
\bibitem [{\citenamefont {Chi}(1999)}]{Chi1999}%
  \BibitemOpen
  \bibfield  {author} {\bibinfo {author} {\bibfnamefont {Z.}~\bibnamefont {Chi}},\ }\bibfield  {title} {\bibinfo {title} {Statistical properties of probabilistic context-free grammars},\ }\href@noop {} {\bibfield  {journal} {\bibinfo  {journal} {Computational Linguistics}\ }\textbf {\bibinfo {volume} {25}},\ \bibinfo {pages} {131} (\bibinfo {year} {1999})}\BibitemShut {NoStop}%
\bibitem [{\citenamefont {Esparza}\ \emph {et~al.}(2013)\citenamefont {Esparza}, \citenamefont {Gaiser},\ and\ \citenamefont {Kiefer}}]{Esparza2013}%
  \BibitemOpen
  \bibfield  {author} {\bibinfo {author} {\bibfnamefont {J.}~\bibnamefont {Esparza}}, \bibinfo {author} {\bibfnamefont {A.}~\bibnamefont {Gaiser}},\ and\ \bibinfo {author} {\bibfnamefont {S.}~\bibnamefont {Kiefer}},\ }\bibfield  {title} {\bibinfo {title} {A strongly polynomial algorithm for criticality of branching processes and consistency of stochastic context-free grammars},\ }\href@noop {} {\bibfield  {journal} {\bibinfo  {journal} {Inf. Process. Lett.}\ }\textbf {\bibinfo {volume} {113}},\ \bibinfo {pages} {381} (\bibinfo {year} {2013})}\BibitemShut {NoStop}%
\bibitem [{\citenamefont {Shieber}(1985)}]{Shieber85}%
  \BibitemOpen
  \bibfield  {author} {\bibinfo {author} {\bibfnamefont {S.~M.}\ \bibnamefont {Shieber}},\ }\bibfield  {title} {\bibinfo {title} {Evidence against the context-freeness of natural language},\ }\href@noop {} {\bibfield  {journal} {\bibinfo  {journal} {Linguist.\ Philos.}\ }\textbf {\bibinfo {volume} {8}},\ \bibinfo {pages} {333} (\bibinfo {year} {1985})}\BibitemShut {NoStop}%
\bibitem [{\citenamefont {Culy}(1985)}]{Culy85}%
  \BibitemOpen
  \bibfield  {author} {\bibinfo {author} {\bibfnamefont {C.}~\bibnamefont {Culy}},\ }\bibfield  {title} {\bibinfo {title} {The complexity of the vocabulary of bambara},\ }\href@noop {} {\bibfield  {journal} {\bibinfo  {journal} {Linguist.\ Philos.}\ }\textbf {\bibinfo {volume} {8}},\ \bibinfo {pages} {345} (\bibinfo {year} {1985})}\BibitemShut {NoStop}%
\bibitem [{\citenamefont {Johnson}\ \emph {et~al.}(2006)\citenamefont {Johnson}, \citenamefont {Griffiths},\ and\ \citenamefont {Goldwater}}]{Johnson2006}%
  \BibitemOpen
  \bibfield  {author} {\bibinfo {author} {\bibfnamefont {M.}~\bibnamefont {Johnson}}, \bibinfo {author} {\bibfnamefont {T.}~\bibnamefont {Griffiths}},\ and\ \bibinfo {author} {\bibfnamefont {S.}~\bibnamefont {Goldwater}},\ }\bibfield  {title} {\bibinfo {title} {Adaptor grammars: A framework for specifying compositional nonparametric bayesian models},\ }\href@noop {} {\bibfield  {journal} {\bibinfo  {journal} {Adv. Neural Inf. Process. Syst.}\ }\textbf {\bibinfo {volume} {19}} (\bibinfo {year} {2006})}\BibitemShut {NoStop}%
\bibitem [{\citenamefont {O'Donnell}\ \emph {et~al.}(2009)\citenamefont {O'Donnell}, \citenamefont {Tenenbaum},\ and\ \citenamefont {Goodman}}]{ODonnell2009}%
  \BibitemOpen
  \bibfield  {author} {\bibinfo {author} {\bibfnamefont {T.~J.}\ \bibnamefont {O'Donnell}}, \bibinfo {author} {\bibfnamefont {J.~B.}\ \bibnamefont {Tenenbaum}},\ and\ \bibinfo {author} {\bibfnamefont {N.~D.}\ \bibnamefont {Goodman}},\ }\href@noop {} {\emph {\bibinfo {title} {Fragment Grammars: Exploring Computation and Reuse in Language}}},\ \bibinfo {type} {Tech. Rep.}\ \bibinfo {number} {MIT-CSAIL-TR-2009-013}\ (\bibinfo  {institution} {Massachusetts Institute of Technology},\ \bibinfo {address} {Cambridge, MA},\ \bibinfo {year} {2009})\BibitemShut {NoStop}%
\bibitem [{\citenamefont {Chomsky}(1956)}]{Chomsky_three}%
  \BibitemOpen
  \bibfield  {author} {\bibinfo {author} {\bibfnamefont {N.}~\bibnamefont {Chomsky}},\ }\bibfield  {title} {\bibinfo {title} {Three models for the description of language},\ }\href@noop {} {\bibfield  {journal} {\bibinfo  {journal} {IRE Transactions on {I}nformation {T}heor.}\ }\textbf {\bibinfo {volume} {2}},\ \bibinfo {pages} {113} (\bibinfo {year} {1956})}\BibitemShut {NoStop}%
\bibitem [{\citenamefont {DeGiuli}(2019{\natexlab{a}})}]{DeGiuli19a}%
  \BibitemOpen
  \bibfield  {author} {\bibinfo {author} {\bibfnamefont {E.}~\bibnamefont {DeGiuli}},\ }\bibfield  {title} {\bibinfo {title} {Random language model},\ }\href@noop {} {\bibfield  {journal} {\bibinfo  {journal} {Phys.\ Rev.\ Lett.}\ }\textbf {\bibinfo {volume} {112}},\ \bibinfo {pages} {128301} (\bibinfo {year} {2019}{\natexlab{a}})}\BibitemShut {NoStop}%
\bibitem [{\citenamefont {DeGiuli}(2019{\natexlab{b}})}]{DeGiuli19b}%
  \BibitemOpen
  \bibfield  {author} {\bibinfo {author} {\bibfnamefont {E.}~\bibnamefont {DeGiuli}},\ }\bibfield  {title} {\bibinfo {title} {Emergence of order in random languages},\ }\href@noop {} {\bibfield  {journal} {\bibinfo  {journal} {J.\ Phys.\ A}\ }\textbf {\bibinfo {volume} {52}},\ \bibinfo {pages} {504001} (\bibinfo {year} {2019}{\natexlab{b}})}\BibitemShut {NoStop}%
\bibitem [{\citenamefont {Hopcroft}\ \emph {et~al.}(2007)\citenamefont {Hopcroft}, \citenamefont {Motwani},\ and\ \citenamefont {Ullman}}]{Hopcroft07}%
  \BibitemOpen
  \bibfield  {author} {\bibinfo {author} {\bibfnamefont {J.~E.}\ \bibnamefont {Hopcroft}}, \bibinfo {author} {\bibfnamefont {R.}~\bibnamefont {Motwani}},\ and\ \bibinfo {author} {\bibfnamefont {J.}~\bibnamefont {Ullman}},\ }\href@noop {} {\emph {\bibinfo {title} {Introduction to Automata Theory,\ Languages,\ and Computation}}},\ \bibinfo {edition} {3rd}\ ed.\ (\bibinfo  {publisher} {Addison-Wesley},\ \bibinfo {address} {Boston},\ \bibinfo {year} {2007})\BibitemShut {NoStop}%
\bibitem [{\citenamefont {J{\"a}ger}\ and\ \citenamefont {Rogers}(2012)}]{jager2012formal}%
  \BibitemOpen
  \bibfield  {author} {\bibinfo {author} {\bibfnamefont {G.}~\bibnamefont {J{\"a}ger}}\ and\ \bibinfo {author} {\bibfnamefont {J.}~\bibnamefont {Rogers}},\ }\bibfield  {title} {\bibinfo {title} {Formal language theory: refining the chomsky hierarchy},\ }\href@noop {} {\bibfield  {journal} {\bibinfo  {journal} {Philosophical Transactions of the Royal Society B: Biological Sciences}\ }\textbf {\bibinfo {volume} {367}},\ \bibinfo {pages} {1956} (\bibinfo {year} {2012})}\BibitemShut {NoStop}%
\bibitem [{Note1()}]{Note1}%
  \BibitemOpen
  \bibinfo {note} {Although the process of rewriting a symbol as a function of the symbol and its neighbors is similar to that of an elementary cellular automaton, our model has several features. In our model, rewriting operations are asynchronous and random. Furthermore, the number of cells or symbols grows exponentially because a single symbol becomes two symbols. A cellular automaton with asynchronous and random updates is called an asynchronous cellular automaton \cite {Fates2018}. Our PCSG can be regarded as a modified version of an asynchronous cellular automaton such that the system size grows.}\BibitemShut {Stop}%
\bibitem [{\citenamefont {Wu}(1982)}]{Wu1982}%
  \BibitemOpen
  \bibfield  {author} {\bibinfo {author} {\bibfnamefont {F.~Y.}\ \bibnamefont {Wu}},\ }\bibfield  {title} {\bibinfo {title} {The {P}otts model},\ }\href@noop {} {\bibfield  {journal} {\bibinfo  {journal} {Rev. Mod. Phys.}\ }\textbf {\bibinfo {volume} {54}},\ \bibinfo {pages} {235} (\bibinfo {year} {1982})}\BibitemShut {NoStop}%
\bibitem [{\citenamefont {Binder}(1981)}]{Binder81}%
  \BibitemOpen
  \bibfield  {author} {\bibinfo {author} {\bibfnamefont {K.}~\bibnamefont {Binder}},\ }\bibfield  {title} {\bibinfo {title} {Finite size scaling analysis of ising model block distribution functions},\ }\href@noop {} {\bibfield  {journal} {\bibinfo  {journal} {Z. Phys. B}\ }\textbf {\bibinfo {volume} {43}},\ \bibinfo {pages} {119} (\bibinfo {year} {1981})}\BibitemShut {NoStop}%
\bibitem [{\citenamefont {Binder}\ and\ \citenamefont {Landau}(1984)}]{Binder84}%
  \BibitemOpen
  \bibfield  {author} {\bibinfo {author} {\bibfnamefont {K.}~\bibnamefont {Binder}}\ and\ \bibinfo {author} {\bibfnamefont {D.~P.}\ \bibnamefont {Landau}},\ }\bibfield  {title} {\bibinfo {title} {Finite-size scaling at first-order phase transitions},\ }\href@noop {} {\bibfield  {journal} {\bibinfo  {journal} {Phys.\ Rev.\ B}\ }\textbf {\bibinfo {volume} {30}},\ \bibinfo {pages} {1477} (\bibinfo {year} {1984})}\BibitemShut {NoStop}%
\bibitem [{\citenamefont {Efron}\ and\ \citenamefont {Tibshirani}(1993)}]{Efron1993}%
  \BibitemOpen
  \bibfield  {author} {\bibinfo {author} {\bibfnamefont {B.}~\bibnamefont {Efron}}\ and\ \bibinfo {author} {\bibfnamefont {R.~J.}\ \bibnamefont {Tibshirani}},\ }\href@noop {} {\emph {\bibinfo {title} {{An Introduction to the Bootstrap}}}},\ Chapman \& Hall/CRC Monographs on Statistics and Applied Probability\ (\bibinfo  {publisher} {Chapman and Hall},\ \bibinfo {address} {London},\ \bibinfo {year} {1993})\BibitemShut {NoStop}%
\bibitem [{\citenamefont {Young}(2012)}]{Young2012}%
  \BibitemOpen
  \bibfield  {author} {\bibinfo {author} {\bibfnamefont {P.}~\bibnamefont {Young}},\ }\bibfield  {title} {\bibinfo {title} {Everything you wanted to know about data analysis and fitting but were afraid to ask},\ }\href@noop {} {\bibfield  {journal} {\bibinfo  {journal} {arXiv:1210.3781}\ } (\bibinfo {year} {2012})}\BibitemShut {NoStop}%
\bibitem [{\citenamefont {Li}(1990)}]{Li1990}%
  \BibitemOpen
  \bibfield  {author} {\bibinfo {author} {\bibfnamefont {W.}~\bibnamefont {Li}},\ }\bibfield  {title} {\bibinfo {title} {Mutual information functions versus correlation functions},\ }\href@noop {} {\bibfield  {journal} {\bibinfo  {journal} {J. Stat. Phys.}\ }\textbf {\bibinfo {volume} {60}},\ \bibinfo {pages} {823} (\bibinfo {year} {1990})}\BibitemShut {NoStop}%
\bibitem [{\citenamefont {Li}(1989{\natexlab{b}})}]{Li1989language}%
  \BibitemOpen
  \bibfield  {author} {\bibinfo {author} {\bibfnamefont {W.}~\bibnamefont {Li}},\ }\bibfield  {title} {\bibinfo {title} {Mutual information functions of natural language texts}} (\bibinfo {year} {1989}{\natexlab{b}}),\ \bibinfo {note} {{SFI} Working Paper, \url{https://www.santafe.edu/research/results/working-papers/mutual-information-functions-of-natural-language-t}}\BibitemShut {NoStop}%
\bibitem [{\citenamefont {Tanaka-Ishii}(2021)}]{Tanaka2021}%
  \BibitemOpen
  \bibfield  {author} {\bibinfo {author} {\bibfnamefont {K.}~\bibnamefont {Tanaka-Ishii}},\ }\href@noop {} {\emph {\bibinfo {title} {Statistical Universals of Language: Mathematical Chance vs. Human choice}}}\ (\bibinfo  {publisher} {Springer},\ \bibinfo {address} {Cham},\ \bibinfo {year} {2021})\BibitemShut {NoStop}%
\bibitem [{\citenamefont {Sainburg}\ \emph {et~al.}(2019)\citenamefont {Sainburg}, \citenamefont {Theilman}, \citenamefont {Thielk},\ and\ \citenamefont {Gentner}}]{Sainburg2019}%
  \BibitemOpen
  \bibfield  {author} {\bibinfo {author} {\bibfnamefont {T.}~\bibnamefont {Sainburg}}, \bibinfo {author} {\bibfnamefont {B.}~\bibnamefont {Theilman}}, \bibinfo {author} {\bibfnamefont {M.}~\bibnamefont {Thielk}},\ and\ \bibinfo {author} {\bibfnamefont {T.~Q.}\ \bibnamefont {Gentner}},\ }\bibfield  {title} {\bibinfo {title} {Parallels in the sequential organization of birdsong and human speech},\ }\href@noop {} {\bibfield  {journal} {\bibinfo  {journal} {Nature {C}ommunications}\ }\textbf {\bibinfo {volume} {10}},\ \bibinfo {pages} {3636} (\bibinfo {year} {2019})}\BibitemShut {NoStop}%
\bibitem [{\citenamefont {Li}\ and\ \citenamefont {Kaneko}(1992)}]{Li1992}%
  \BibitemOpen
  \bibfield  {author} {\bibinfo {author} {\bibfnamefont {W.}~\bibnamefont {Li}}\ and\ \bibinfo {author} {\bibfnamefont {K.}~\bibnamefont {Kaneko}},\ }\bibfield  {title} {\bibinfo {title} {{Long-Range} correlation and partial 1/f spectrum in a noncoding {DNA} sequence},\ }\href@noop {} {\bibfield  {journal} {\bibinfo  {journal} {Europhys.\ Lett.}\ }\textbf {\bibinfo {volume} {17}},\ \bibinfo {pages} {655} (\bibinfo {year} {1992})}\BibitemShut {NoStop}%
\bibitem [{\citenamefont {Grassberger}(2003)}]{Grassberger2003}%
  \BibitemOpen
  \bibfield  {author} {\bibinfo {author} {\bibfnamefont {P.}~\bibnamefont {Grassberger}},\ }\bibfield  {title} {\bibinfo {title} {Entropy estimates from insufficient samplings},\ }\href@noop {} {\bibfield  {journal} {\bibinfo  {journal} {arXiv preprint physics/0307138}\ } (\bibinfo {year} {2003})}\BibitemShut {NoStop}%
\bibitem [{\citenamefont {Fukui}(2015)}]{Fukui2015}%
  \BibitemOpen
  \bibfield  {author} {\bibinfo {author} {\bibfnamefont {N.}~\bibnamefont {Fukui}},\ }\bibfield  {title} {\bibinfo {title} {A note on weak vs. strong generation in human language},\ }\href@noop {} {\bibfield  {journal} {\bibinfo  {journal} {Studies in Chinese Linguistics}\ }\textbf {\bibinfo {volume} {36}},\ \bibinfo {pages} {59} (\bibinfo {year} {2015})}\BibitemShut {NoStop}%
\bibitem [{\citenamefont {Joshi}(1985)}]{Joshi1985}%
  \BibitemOpen
  \bibfield  {author} {\bibinfo {author} {\bibfnamefont {A.~K.}\ \bibnamefont {Joshi}},\ }\bibinfo {title} {Tree adjoining grammars: How much context-sensitivity is required to provide reasonable structural descriptions?},\ in\ \href@noop {} {\emph {\bibinfo {booktitle} {Natural Language Parsing: Psychological, Computational, and Theoretical Perspectives}}},\ \bibinfo {series and number} {Studies in Natural Language Processing},\ \bibinfo {editor} {edited by\ \bibinfo {editor} {\bibfnamefont {D.~R.}\ \bibnamefont {Dowty}}, \bibinfo {editor} {\bibfnamefont {L.}~\bibnamefont {Karttunen}},\ and\ \bibinfo {editor} {\bibfnamefont {A.~M.}\ \bibnamefont {Zwicky}}}\ (\bibinfo  {publisher} {Cambridge University Press},\ \bibinfo {year} {1985})\ pp.\ \bibinfo {pages} {206--250}\BibitemShut {NoStop}%
\bibitem [{\citenamefont {Steedman}(1987)}]{Steedman1987}%
  \BibitemOpen
  \bibfield  {author} {\bibinfo {author} {\bibfnamefont {M.}~\bibnamefont {Steedman}},\ }\bibfield  {title} {\bibinfo {title} {Combinatory grammars and parasitic gaps},\ }\href@noop {} {\bibfield  {journal} {\bibinfo  {journal} {Natural Language \& Linguistic Theory}\ }\textbf {\bibinfo {volume} {5}},\ \bibinfo {pages} {403} (\bibinfo {year} {1987})}\BibitemShut {NoStop}%
\bibitem [{\citenamefont {Stabler}(1997)}]{Stabler1997}%
  \BibitemOpen
  \bibfield  {author} {\bibinfo {author} {\bibfnamefont {E.}~\bibnamefont {Stabler}},\ }\bibfield  {title} {\bibinfo {title} {Derivational minimalism},\ }in\ \href@noop {} {\emph {\bibinfo {booktitle} {Logical Aspects of Computational Linguistics}}},\ \bibinfo {editor} {edited by\ \bibinfo {editor} {\bibfnamefont {C.}~\bibnamefont {Retor{\'e}}}}\ (\bibinfo  {publisher} {Springer},\ \bibinfo {address} {Berlin},\ \bibinfo {year} {1997})\ pp.\ \bibinfo {pages} {68--95}\BibitemShut {NoStop}%
\bibitem [{\citenamefont {Fat{\`e}s}(2018)}]{Fates2018}%
  \BibitemOpen
  \bibfield  {author} {\bibinfo {author} {\bibfnamefont {N.}~\bibnamefont {Fat{\`e}s}},\ }\bibfield  {title} {\bibinfo {title} {Asynchronous cellular automata},\ }in\ \href@noop {} {\emph {\bibinfo {booktitle} {Encyclopedia of Complexity and Systems Science}}}\ (\bibinfo  {publisher} {Springer},\ \bibinfo {address} {New York},\ \bibinfo {year} {2018})\ p.~\bibinfo {pages} {21}\BibitemShut {NoStop}%
\end{thebibliography}%

\end{document}


\preprint{APS/123-QED}

\title{Supplemental Material for ``Statistical properties of probabilistic context-sensitive grammars''}

\author{Kai Nakaishi}
\affiliation{
    Graduate School of Arts and Sciences,
    The University of Tokyo,
    Komaba, Meguro-ku, Tokyo 153-8902, Japan
}
\author{Koji Hukushima}
\affiliation{
    Graduate School of Arts and Sciences,
    The University of Tokyo,
    Komaba, Meguro-ku, Tokyo 153-8902, Japan
}
\affiliation{
    Komaba Institute for Science, The
    University of Tokyo, 3-8-1 Komaba, Meguro-ku, Tokyo 153-8902, Japan
}
\date{\today}



\maketitle
In this supplemental material, we present detailed descriptions of the convergence of the distance between the distribution of symbols in the context-free case $q = 0$ and that in the context-sensitive case $q > 0$, the dependence of mutual information between two nodes on the parameter $\epsilon$ and context sensitivity $q$, and the dependence of context-free independence breaking on $\epsilon$ and $q$. These are presented in Sec.~\ref{sec_convergence}, \ref{sec_MI}, and \ref{sec_CMI}, respectively.

\section{Convergence of $\Delta$}
\label{sec_convergence}

\begin{figure*}[h]
    \centering
    \includegraphics[width = 1 \linewidth]{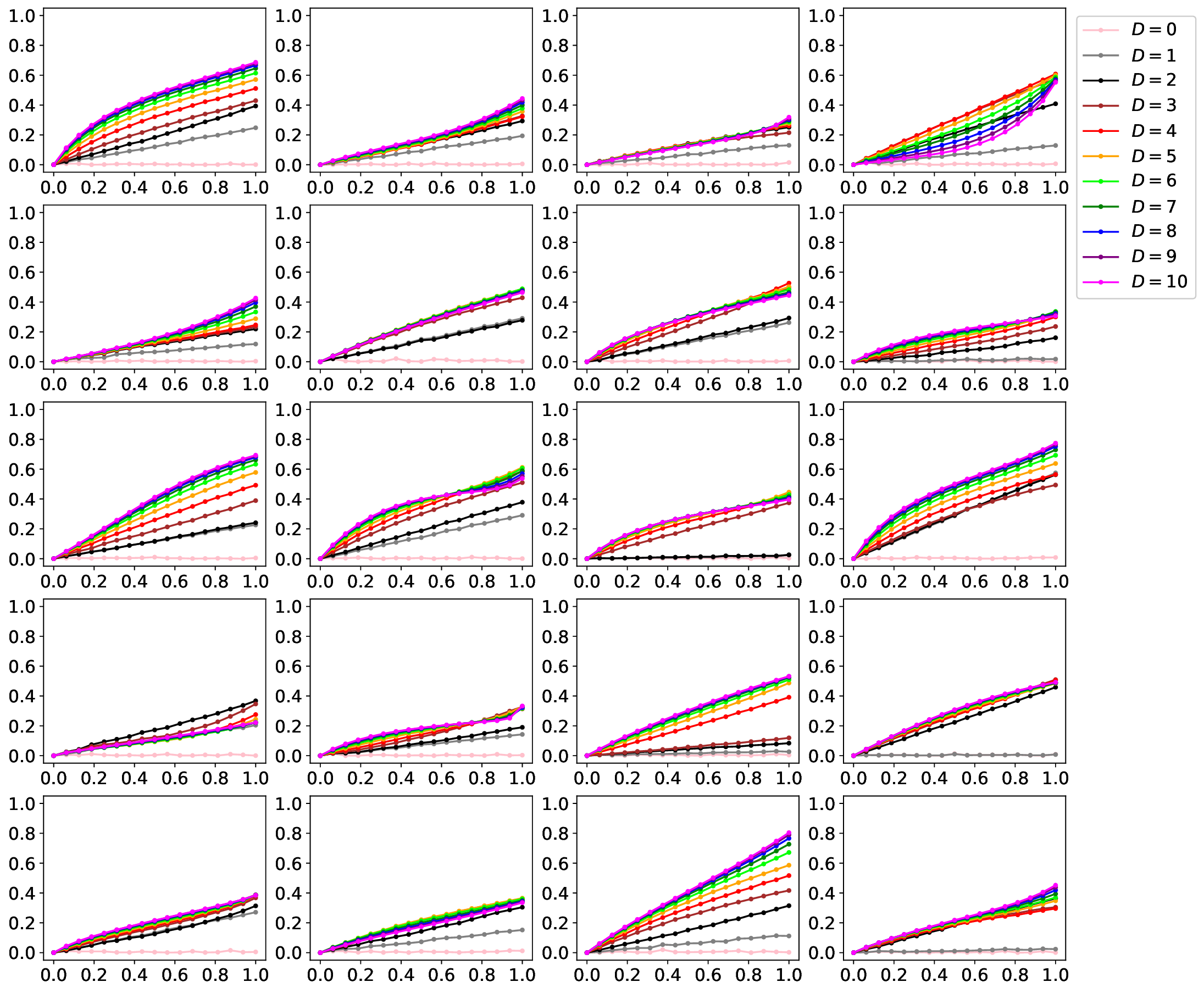}
    \caption{\label{fig_SM_DP} Distances $\Delta (D, q, M)$ between $\pi_{A, i}$ with $q = 0$ and that with $q > 0$ as functions of $q$, computed from sampled $10^4$ trees. Each subfigure presents the $\Delta$s for each $M$ with depth $D = 0, 1, \cdots, 10$. Different colors represent different depths.}
\end{figure*}

In Sec.~III of the main paper, we discussed how the distribution of symbols on nodes changes as a function of the context sensitivity. We noted that the distance $\Delta$ between the distribution in the context-free case $q = 0$ and that in the context-sensitive case $q > 0$ nearly converges when the tree depth is $D = 10$. Figure~\ref{fig_SM_DP} shows $\Delta$s for different depth $D$, as a function of the context sensitivity $q$ for $20$ different $M$s sampled with fixed $\epsilon = 10^{- 2}$. Each subfigure corresponds to each $M$, whereas each color corresponds to each $D$ from $D=1$ to $10$. For most of the $M$s, the $\Delta$s seem to almost converge at $D = 10$. Some of them oscillate, but they depend only moderately on $q$, showing no indication of convergence to any non-analytic function of $q$.

\section{Dependence of the mutual information on $\epsilon$ and $q$}
\label{sec_MI}

In Sec.~IV of the main paper, we examined the mutual information between two nodes. Particularly, we discussed the decay in the mutual information with respect to the effective distance, showing results for the parameter $\epsilon=10^{-2}$ and the context sensitivity $q=1$. At the end of the section, we also provided a brief discussion of the dependence of the mutual information on $\epsilon$ and $q$. Here, we present additional results for different $\epsilon$ and $q$ to complement the discussion.

\begin{figure*}[h]
    \centering
    \includegraphics[width = 1 \linewidth]{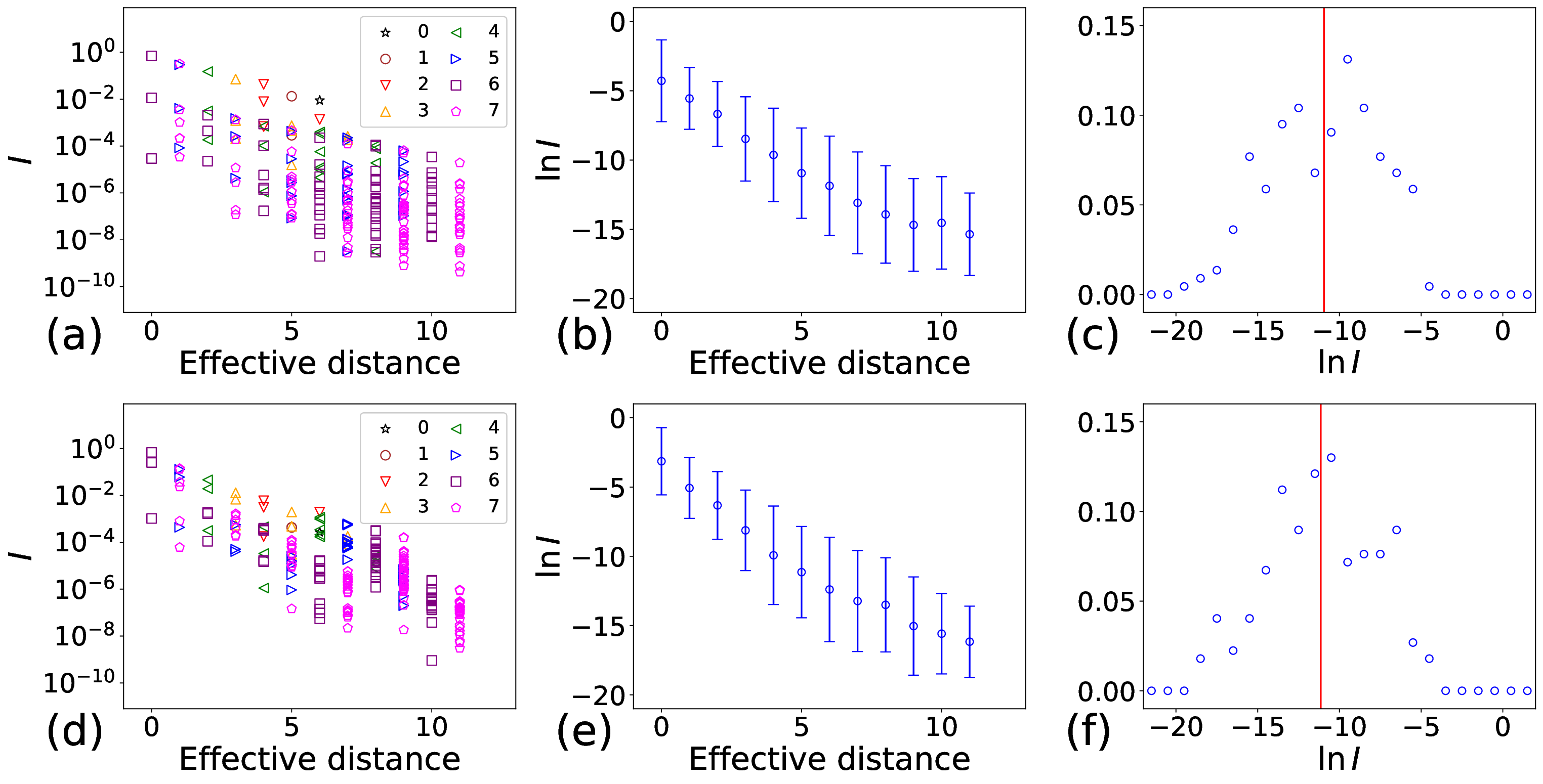}
    \caption{\label{fig_SM_MI_effective}(a) Mutual information $I$ against the effective distance between $i$ and $j$.
        Weights $M$ are generated from the lognormal distribution.
        Markers and colors differ for different depths.
        (b) Averages and standard deviations of $\ln I$ over $j$s of the same effective distance and over 10 $M$s generated.
        (c) Normalized histograms of $\ln I$ for 10 $M$s for which the effective distance $5$.
        The red vertical line represents the average.
        For all (a), (b), and (c), the parameter in the lognormal distribution is $\epsilon = 10^{-2}$, the context sensitivity is $q = 0.5$, and $i$ is fixed at $(1, 0, 0, 0, 0, 0)$.
        For (d), (e), and (f), the results are shown for the same setting except $\epsilon = 10^{-1}$ and $q = 1$.}
\end{figure*}

Figures~\ref{fig_SM_MI_effective}(a), (b), and (c) present the mutual information under the same setting as in Fig.~8 in Sec.~IV of the main paper, except for the context sensitivity $q = 0.5$, which is smaller than $q = 1$ in the main paper. The mutual information $I$ of a particular $M$ for $q = 0.5$, as shown in Fig.~\ref{fig_SM_MI_effective}(a), decays exponentially with the effective distance with a larger rate or smaller correlation length compared to those for $q = 1$. This means that the correlation between nodes tends to be weaker with smaller context sensitivity because each iteration refers to the context with a lower probability. The mutual information averaged over nodes and $M$s, as shown in Fig.~\ref{fig_SM_MI_effective} (b), suggests that this tendency is common to the randomly generated $M$s. The histogram of the mutual information for effective distance $5$ presented in Fig.~\ref{fig_SM_MI_effective} (c) shows that the mutual information is distributed around the average. The setting in Figs.~\ref{fig_SM_MI_effective}(d), (e), and (f) is also the same, except that the parameter of the lognormal distribution of $M$ is set to $\epsilon = 10^{-1}$, which is larger than $\epsilon = 10^{-2}$ in the main paper. The result is similar to the case with $q = 0.5$. The mutual information for $\epsilon = 10^{-1}$ decays faster than for $\epsilon = 10^{- 2}$.  This is also reasonable because a larger $\epsilon$ implies that $M_{ABC}$s are closer to $1/2^2$, implying the uniform distribution of a tree. Under such a PCSG, the structures of the trees are more disordered with less correlation between nodes. We have confirmed that this tendency for the decay rates in $I$ holds for several other $q$s and $\epsilon$s.

\section{Dependence of the context-free independence breaking on $\epsilon$ and $q$}
\label{sec_CMI}

In Sec.~V of the main paper, we proposed the mutual information between the children of two symbol-fixed nodes as a metric to quantify the context-free independence breaking, denoted by $J$. As noted at the end of the section, the dependence of $J$ on $\epsilon$ and $q$ is similar to that of $I$. In this section, we present a concrete explanation by showing numerical results.

\begin{figure*}[h]
    \centering
    \includegraphics[width = 1 \linewidth]{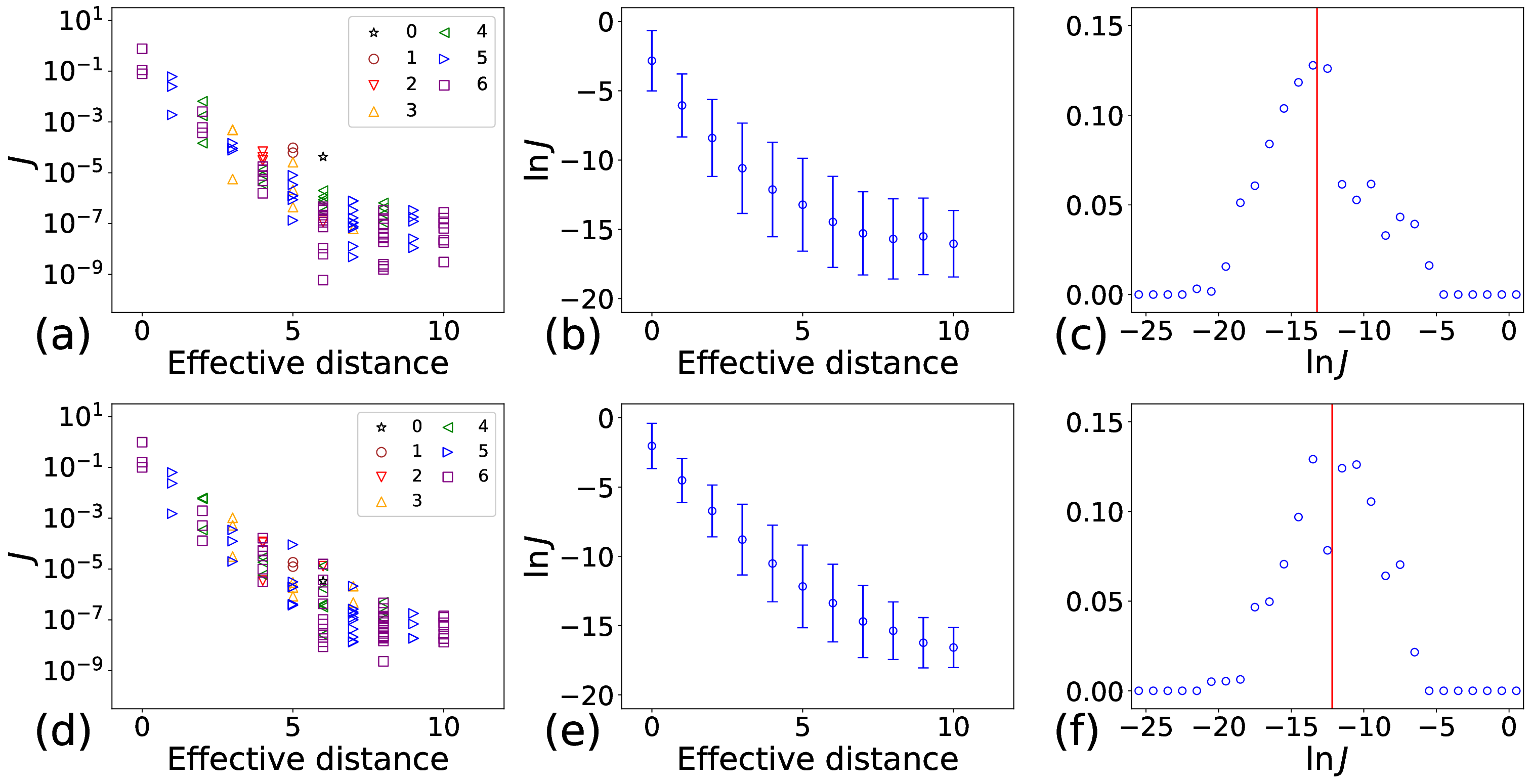}
    \caption{\label{fig_SM_CMI_effective}(a) Degree of the context-free independence breaking, or parent-fixed mutual information $J$ for $A = B = 0$ against the effective distance between $i$ and $j$.
        Weights $M$ are generated from the lognormal distribution.
        Markers and colors differ for different depths.
        (b) Averages and standard deviations of $\ln J$ over $j$s of the same effective distance, over the symbols $A$ of node $i$ and $B$ of $j$, and over 10 $M$s generated.
        (c) Normalized histograms of $\ln J$ for 10 $M$s with effective distance 5.
        The red vertical line represents the average.
        For all (a), (b), and (c), the parameter in the lognormal distribution is $\epsilon = 10^{-2}$, the context sensitivity is $q = 0.5$, and $i$ is fixed at $(1, 0, 0, 0, 0, 0)$.
        (d), (e), and (f) present the results for the same setting except $\epsilon = 10^{-1}$ and $q = 1$.}
\end{figure*}

The effective distance dependence of $J$s for a given $M$ and that of $J$ averaged over nodes, over symbols of the nodes, and over $M$s are shown in Fig.~\ref{fig_SM_CMI_effective}, as well as the distribution of the $J$s for depth $5$, as in Fig.~10 in Sec.~V of the main paper. In the main paper, the parameters were set to $q=1$ and $\epsilon=10^{-2}$. In this section, $q=0.5$ and $\epsilon=10^{-2}$ in (a), (b), and (c),  whereas $q=1$ and $\epsilon = 10^{- 1}$ in (d), (e) and (f). As with the results of $I$ in the preceding section, the decay rates in both cases are larger than those in the main paper. The results with the lower context sensitivity are straightforward to comprehend since $J$ quantifies the context-free independence breaking. The interpretation of the results with larger $\epsilon$ is the same as for $I$. We have confirmed these observations for several other $q$s and $\epsilon$s. Repeatedly, it is non-trivial that the behavior of $I$ and $J$ is qualitatively similar. The bias due to discarding non-positive $J$s is slightly pronounced
because of the small values of $J$ and limited data size.

